\documentclass[reprint, amsmath,amssymb,prb,aps]{revtex4-1}

\usepackage{braket} 
\usepackage{graphicx}
\usepackage{dcolumn}
\usepackage{bm}
\usepackage{color}
\usepackage{lipsum}

\begin{document}

\title{Stimulated Emission Tomography: Beyond Polarization}

\author{Mario Arnolfo Ciampini, Andrea Geraldi,Valeria Cimini }%
 \affiliation{Dipartimento di Fisica, Sapienza University of Rome, Piazzale Aldo Moro 5, 00185 Rome, Italy}
 \author{Chiara Macchiavello}
   \affiliation{Dipartimento di Fisica, University of Pavia, via Bassi 6, 27100 Pavia, Italy} 
   \affiliation{INFN Sezione di Pavia, via Bassi 6, I-27100, Pavia, Italy}
   \affiliation{CNR-INO, largo E. Fermi 6, I-50125, Firenze, Italy }
  \author{J.E. Sipe}
  \affiliation{Department of Physics, University of Toronto, 60 St. George St., Toronto, ON, M5S 1A7, Canada}
\author{M. Liscidini}
 \email{marco.liscidini@unipv.it}
\affiliation{Dipartimento di Fisica, Universit\`a degli studi di Pavia, Via Bassi 6, 27100 Pavia, Italy}
\author{Paolo Mataloni}
 \affiliation{Dipartimento di Fisica, Sapienza University of Roma, Piazzale Aldo Moro 5, 00185 Rome, Italy}

\begin{abstract}
In this work we demonstrate the use of stimulated emission tomography to characterize a hyper-entangled state generated by spontaneous parametric down-conversion in a CW-pumped source.  In particular, we consider the generation of hyper-entangled states consisting of photon pairs entangled in polarisation and path.  These results extend the capability of stimulated emission tomography beyond the polarisation degree of freedom, and demonstrate the use of this technique to study states in higher dimension Hilbert spaces.    
\end{abstract}

\maketitle
Quantum state tomography (QST) is the task of experimentally identifying the density matrix describing a quantum state \cite{kwiat:tomography}. For quantum optical systems, it involves a number of coincidence measurements, depending on the size of the Hilbert space under investigation. For instance, for two polarization-entangled photons, $2^4$-1=15 independent polarization-resolved coincidence measurements are required \cite{kwiat:tomography}. More  generally, the dimension of the Hilbert space, and thus the number of measurements required, depends on both the number of photons and the degrees of freedom (DOFs) - such as polarization, spatial structure, etc. --  that are involved.

While in the initial work on QST of photonic states only one DOF at a time - typically polarization \cite{white_99} - was considered, in the last fifteen years there has been growing interest in exploiting quantum correlations using different photon DOFs simultaneously. In particular, even states that are entangled in every DOF have been demonstrated by Kwiat et al. \cite{Barreiro_05}, where the state of one photon pair belongs to a 36-dimensional Hilbert space. The generation and manipulation of such states is of paramount importance in quantum optics, for it allows for an increase in the information that can be stored in a state without the need for an increased  number of photons. Yet the QST of these states can be extremely challenging, with  large numbers of difficult coincidence measurements required. 

In the last few years, there has also been progress in the generation and manipulation of non-classical light by exploiting parametric fluorescence in photonic integrated circuits (PICs), using either spontaneous parametric down conversion (SPDC) or spontaneous four-wave mixing (SFWM) \cite{Ducci_14,horn_12,Grassani_15,Silverstone_15,chen_16}. In these systems, light confinement at the micron scale leads to an enhancement of the efficiency of parametric fluorescence over what can be achieved with bulk crystals by up to seven orders of magnitude \cite{atzeni2018integrated}. In the case of PICs the use of the polarization DOF is particularly challenging, and so path and energy are usually the preferred DOFs for quantum correlations. Even here QST is difficult, for the photon collection efficiency is still done off-chip and is plagued by coupling and propagation losses, as well as less-than-ideal detector efficiencies. 

A few years ago, it was suggested that sources based on parametric fluorescence could also be characterized by Stimulated Emission Tomography (SET), which relies on the relation between the spontaneous and stimulated emission of photon pairs \cite{liscidini_13}.  In SET the density matrix that would be relevant in a spontaneous process is determined by characterizing the associated stimulated process. Thus higher signal-to-noise can be achieved than in a spontaneous emission experiment, and the equipment necessary for coincidence measurements (i.e. single-photon detector) is usually not required. The validity of this approach has been demonstrated for polarization-entangled photon pairs generated by a sandwich BBO crystal source \cite{Rozema:15}, with the dependence of the polarization density matrix on the energies of the emitted photons generated reported in \cite{Fang:16} for SFWM in optical fibers. However, so far, only polarization density matrices have been reconstructed via SET experiments. Demonstrating SET on other DOFs than polarization would be an important milestone towards the use of this powerful tool for the characterization of sources of non-classical light in several platforms.

In this work we demonstrate SET on multiple DOFs for the first time, and  determine the reduced density matrix for both polarization and path in the case of a two-photon hyperentagled state. These results are particularly important, not only because the path Hilbert space can have, in principle, infinite dimension, but also because we extend the use of SET to one of the preferred DOFs in PICs \cite{crespi2011integrated,corrielli2014rotated,silverstone_13,Wangeaar7053}.  Indeed, previous characterizations of photon pair sources via stimulated emission have shown that in integrated devices or optical fibers, where there is a discrete set of spatial modes, SET can outperform  traditional methods based on parametric fluorescence in terms of resolution and speed \cite{Eckstein_14,Fang:14,Iman_15,Grassani_16,Fang:16}.  However, here we make a different choice and use SET to study the generation of photon pairs in a bulk nonlinear crystal, in which the realization of entanglement in multiple DOFs is easier. More importantly, we want to compare SET results with those obtained  with QST, allowing for their  better understanding.

We consider a $1.5$ mm long type I crystal of $\beta$-Barium Borate (BBO), excited in two opposite directions by a 100 mW, vertical polarized (V), continuous-wave (cw) pump laser with wavelength $\lambda_{p}=355$ nm (GENESIS, Coherent) (see Fig. \ref{fig:source}). In the first passage through the BBO crystal, pairs of horizontally polarized (H) photons could be emitted and would pass through a wide-band  quarter wave plate (QWP), be back-reflected by a spherical mirror,  and pass again through the same QWP to be vertically polarized. The mirror, with radius of curvature $R=15$ cm, is placed a distance $d=R$ from the BBO crystal. The pump laser is also back-reflected by the same mirror and excites the BBO crystal a second time, with the possibility of creating a pair of horizontally polarized photons. When spatial and temporal overlapping of the two generations is guaranteed by a proper arrangement of the optical elements and the cw pump, respectively, the generated photons are in a quantum superposition. This source is a modified version of that reported earlier \cite{mata:hyperrealization}, in which path-polarization hyperentangled states are generated in an energy-degenerate configuration. However, in the present implementation the photons are generated at  two different wavelengths: $\lambda_{1}\cong 656$ nm and $\lambda_{2}\cong 777$ nm. Since photon pairs are emitted in a spatial conical distribution determined by the SPDC phase-matching condition, path entanglement is expected and can be verified by selecting photons along four directions by using a four-hole mask having the holes  located along the horizontal diameter of the circular section of the emission cone (see Fig. \ref{fig:mask}).

In this simple picture, polarization and path DOFs are independent. Thus one would expect the hyperentangled state 
\begin{equation}\label{nocorr}
\ket{\psi_{\mathrm{hyper}}}=\ket{\psi_{\mathrm{path}}} \otimes\ket{\psi_{\mathrm{pol}}},
\end{equation}
with polarization state 
\begin{equation}
\label{eq:polstate}
\ket{\psi_{\mathrm{pol}}}=\frac{1}{\sqrt{2}}(\ket{H_{\lambda_1}H_{\lambda_2}}+\text{e}^{i\phi}\ket{V_{\lambda_1}V_{\lambda_2}}),
\end{equation}
and path state 
\begin{equation}
\label{eq:pathstate}
\ket{\psi_{\mathrm{path}}}=\frac{1}{\sqrt{2}}(\ket{A_{\lambda_{1}}B_{\lambda_{2}}}+\text{e}^{i\theta}\ket{B_{\lambda_{1}}A_{\lambda_{2}}}),
\end{equation}
where photons are labeled by their wavelength and can exit either through the left (A) or the right (B) hole with respect to the center of the mask. Finally, $\phi$ and $\theta$ are phase factors, with the former depending on the displacement of the spherical mirror along the laser pump direction, and the latter being controlled by means of a phase shifter (PS) consisting of a thin glass plate placed in one of the four spatial modes.

 \begin{figure}[!ht]
 		\centering%
 		\includegraphics[width=0.44\textwidth]{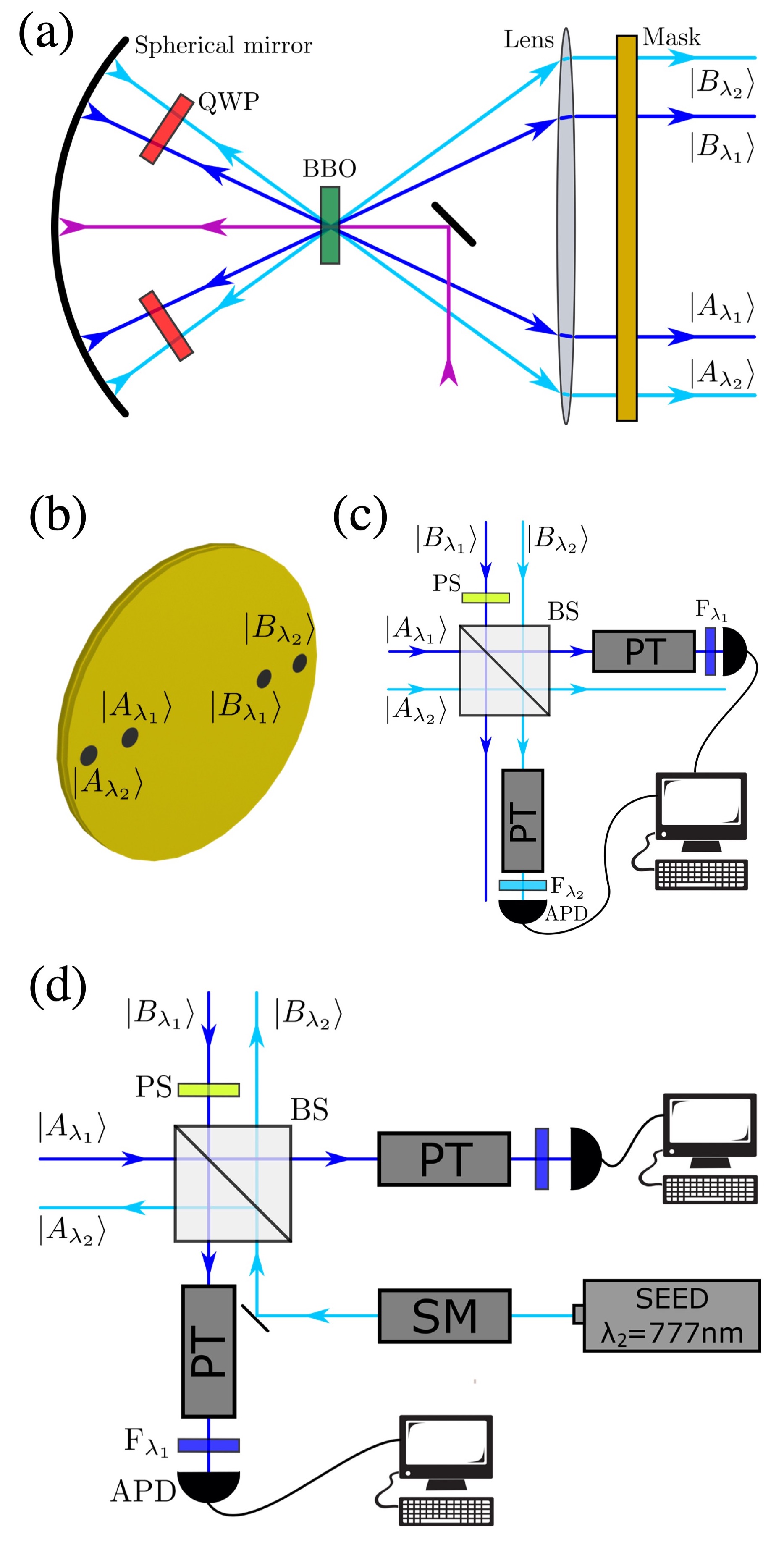}
 		\caption{\color{black}{\emph{Experimental apparatus. (a) Sketch of the path-polarization entangled photons source with non degenerate wavelengths. QWP: Quarter Wave Plates (QWP); BBO: BBO crystal. (b) Sketch of the mask used to select modes for path entanglement. (c) Sketch of the setup used to perform QST. Here the Polarization-Tomography (PT) boxes represent the QWP, HWP (Half Wave Plate) and the PBS (Polarizing Beam Splitter) needed to perform the Pauli measurements. BS: Beam Splitter; PS: Phase Shifter; APD: Avalanches Photo Detector; $F_{\lambda_i}$: interference filters at $\lambda_1$ and $\lambda_2$. (d) Sketch of the setup used to perform SET. Here the Seed-Modifier (SM) box represents the QWP, HWP, BS, and PS needed to modify the seed characteristics.}}}\label{fig:source}
 	\end{figure}

 \begin{center}
 \begin{table*}[!ht]
 \color{black}{
 {\small{}
  \hfill{}
  \begin{tabular}{ccccc}
  \em Parameter & \em Path QST & \em Polarization QST & \em Path SET & \em Polarization SET \\
  \hline
  F & 0.943$\pm$0.002 & 0.857$\pm$0.008 & 0.934$\pm$0.001 & 0.814$\pm$0.008 \\
  \hline
  Tr($\rho^2$) & 0.909$\pm$0.003 & 0.772$\pm$0.014 & 0.886$\pm$0.001 & 0.694$\pm$0.012 \\
  \hline
  $\tau$  & 0.785$\pm$0.005 & 0.577$\pm$0.026 & 0.779$\pm$0.001 & 0.411$\pm$0.022 \\
  \hline
  $C$ & 0.886$\pm$0.003 & 0.759$\pm$0.017 & 0.883$\pm$0.001 & 0.641$\pm$0.017
 \end{tabular}}
 }
 \hfill{}
  \caption{\emph{Relevant parameters derived from the measured density matrices using QST and SET. The trace of the square of the density matrix, the fidelity with the expected matrix, the tangle and the concurrence for both polarization and path DOF's are reported. The fidelity for polarization is computed respect to the state \ref{eq:polstate} with $\phi=0$. In the case of path it is computed with respect to the state \ref{eq:pathstate} with $\theta=0$.}}
 \label{tab:quantumparameters}
 \end{table*}
 \end{center}

\begin{figure}[!ht]
 	\centering
 	\includegraphics[width=0.45\textwidth]{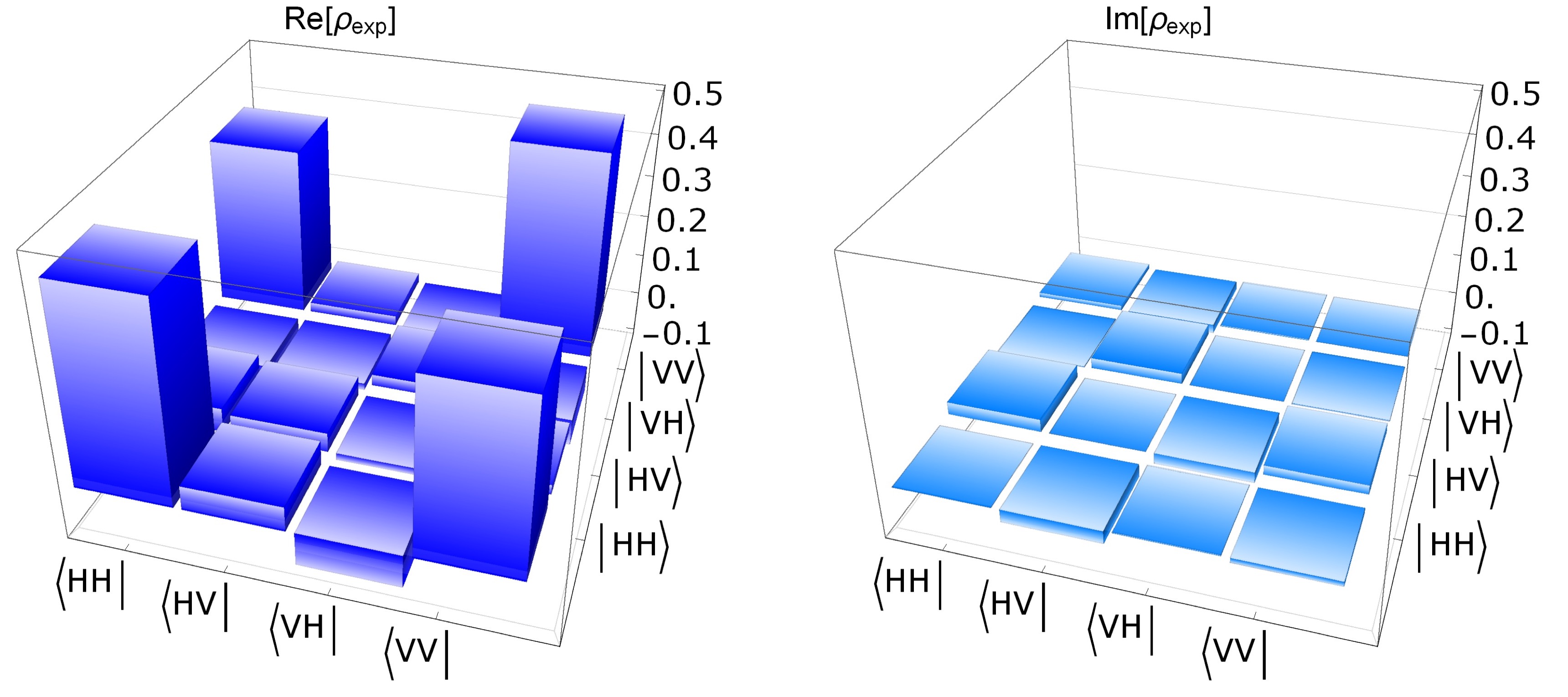}%
 	\caption{\emph{Reconstructed density matrix for polarization degree of freedom using QST: real part (left) and imaginary part (right).}}
 	\label{fig:polarizationquantum}
 \end{figure}

\begin{figure}[!h]
 	\centering
 	\includegraphics[width=0.45\textwidth]{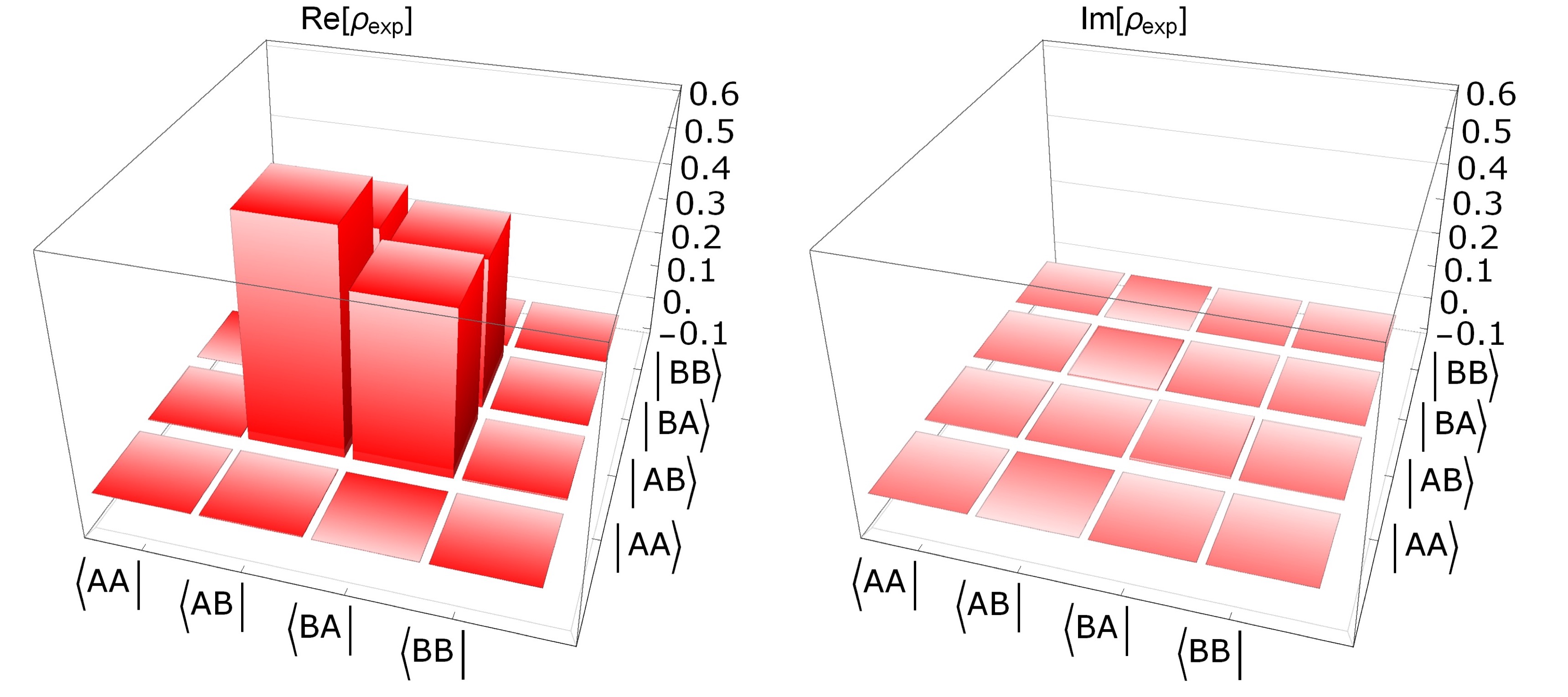}
 	\caption{\emph{Reconstructed density matrix for path degree of freedom using QST: real part (left) and imaginary part (right).}}
 	\label{fig:pathquantum}
 \end{figure}

First, we characterize the generated state via QST for both DOFs  by measuring the mean values of the corresponding Pauli operators $\hat{\sigma}_{x},\hat{\sigma}_{y}$, and $\hat{\sigma}_{z}$. As usual, in the case of polarization, this is done by means of QWPs, half wave plates (HWPs), and polarizing beam splitters (PBS); a PS and a beam splitter (BS) are used to construct the necessary observables in the path DOF (see Fig. \ref{fig:QST}). Finally, single-mode fibers are used to direct photons to two single-photon avalanche photodetectors (APDs), \textcolor{black}{while two $3$nm-bandwidth interference filters ($F_{\lambda_{1/2}}$) are used to separate the photons according to their wavelength $\lambda_1$ and $\lambda_2$. The measured coincidence rate is about $100$ Hz in a gate temporal window  of $9$ ns with a coincidence to accident ratio of the order of 100.} QST required about 10 minutes. The path and polarization density matrices reconstructed using these experimental data by hypercomplete quantum state tomography \cite{kwiat:tomography,kwiat:numericoptimization} are shown in Fig. \ref{fig:polarizationquantum} and Fig. \ref{fig:pathquantum}, respectively.
From the density matrices, we also calculate fidelities, purities, tangles, and concurrences (see Tab. \ref{tab:quantumparameters}.) Uncertainties are computed by assuming the coincidence counts to be Poissonian distributed, and neglecting any systematic error.

 
The path \emph{fidelity} is close to unity, with a small discrepancy that we attribute to the limited visibility of the interferometer used in the QST measurement. In contrast, the polarization fidelity is considerably lower than unity, which indicates that our initial model of the source may be too simplistic. This is confirmed by the \emph{purity} values reported in Tab. 1, where the polarization purity is significantly less than that of the path. 

The difference in purity between path and polarization suggests that one might have to consider additional DOFs for a correct interpretation of the results. In our initial description, we did not take into account that the photons are collected within a certain momentum range associated with the finite aperture of the hole in the mask.  Taking this into account requires the introduction of additional DOF and its corresponding Hilbert space, such that:
\begin{equation}
	\label{hilbert3}
  \mathcal{H}=\mathcal{H_{\mathrm{Path}}}\otimes\mathcal{H}_{\mathrm{\mathrm{Pol}}}\otimes\mathcal{H}_{\kappa},
\end{equation}
where $\mathcal{H}_{\kappa}$ is associated with the momentum DOF determined by the hole size. Thus, the generated state becomes:
\begin{equation}
\label{hyper_k}
\ket{\psi_{\mathrm{hyper}}}=\ket{\psi_{\mathrm{path}}} \otimes\ket{\psi_{\mathrm{pol},\kappa}},
\end{equation}
where 
\begin{equation}
\label{pol_k}
\ket{\psi_{\mathrm{pol},\kappa}}\neq\ket{\psi_{\mathrm{pol}}} \otimes\ket{\psi_\kappa},
\end{equation}
since polarization and momentum are entangled. This would explain the unit purity for the path density operator and a lower purity for the polarization density operator. This description is consistent with previous experimental results \cite{Rozema:15,Fang:16,Altepeter:05}. 

We now turn to the SET measurements, remembering that in our experiment the two generated photons have different energies. This choice is motivated by our use of a cw-pumped source, which results in a lower efficiency than that of sources studied in earlier works \cite{Rozema:15,Fang:16}. In our scenario, the use of a seed with the same wavelength as that of the generated light would lead to a generated signal intensity comparable with that of the seed light scattered by the nonlinear crystal. Thus, to improve the signal-to-noise ratio, it is convenient to work with non-degenerate SPDC, where the seed beam can be filtered before detection. 

\begin{figure}[!ht]
 	\centering
 	\includegraphics[width=0.45\textwidth]{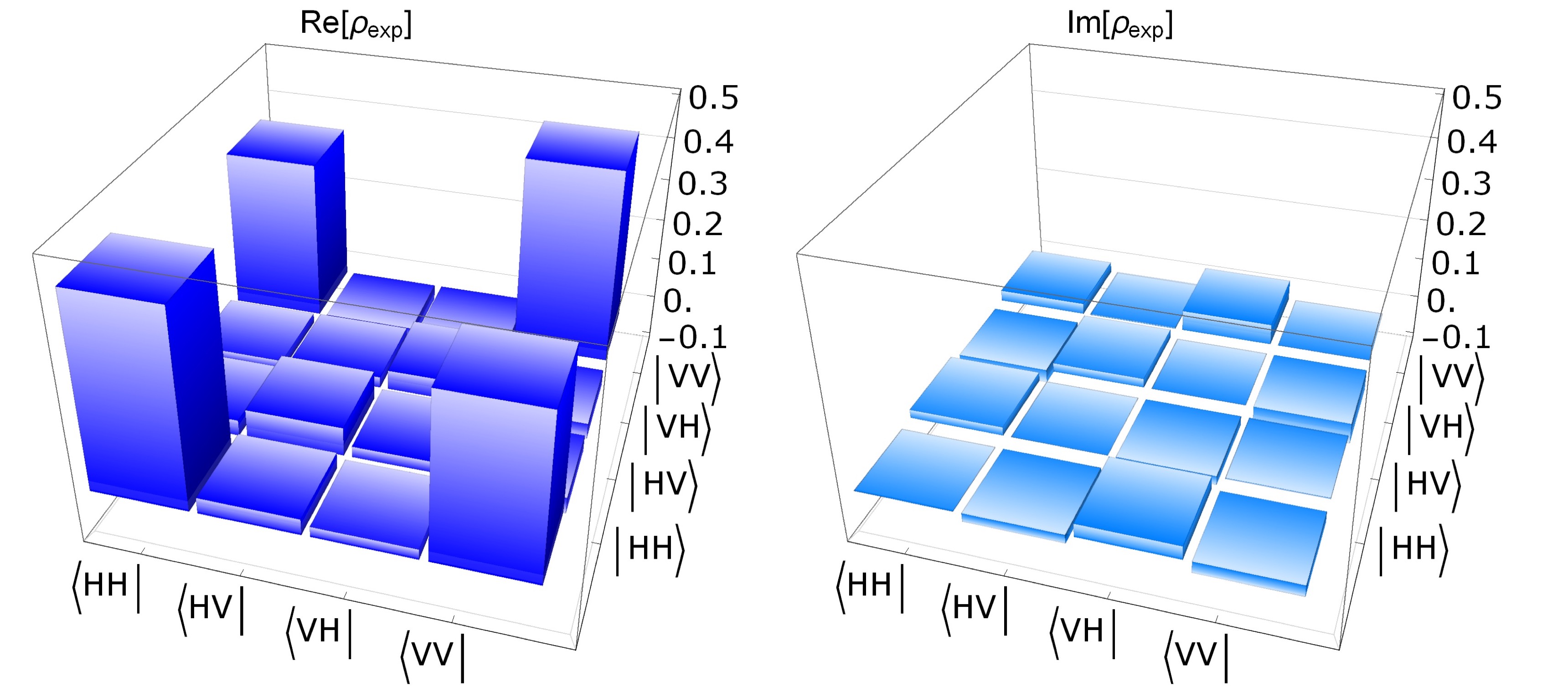}
 	\caption{\emph{Reconstructed density matrix for polarization degree of freedom using SET: real part (left) and imaginary part (right).}}
 	\label{fig:polarizationSET}
 \end{figure}

\begin{figure}[!ht]
 	\centering
 	\includegraphics[width=0.45\textwidth]{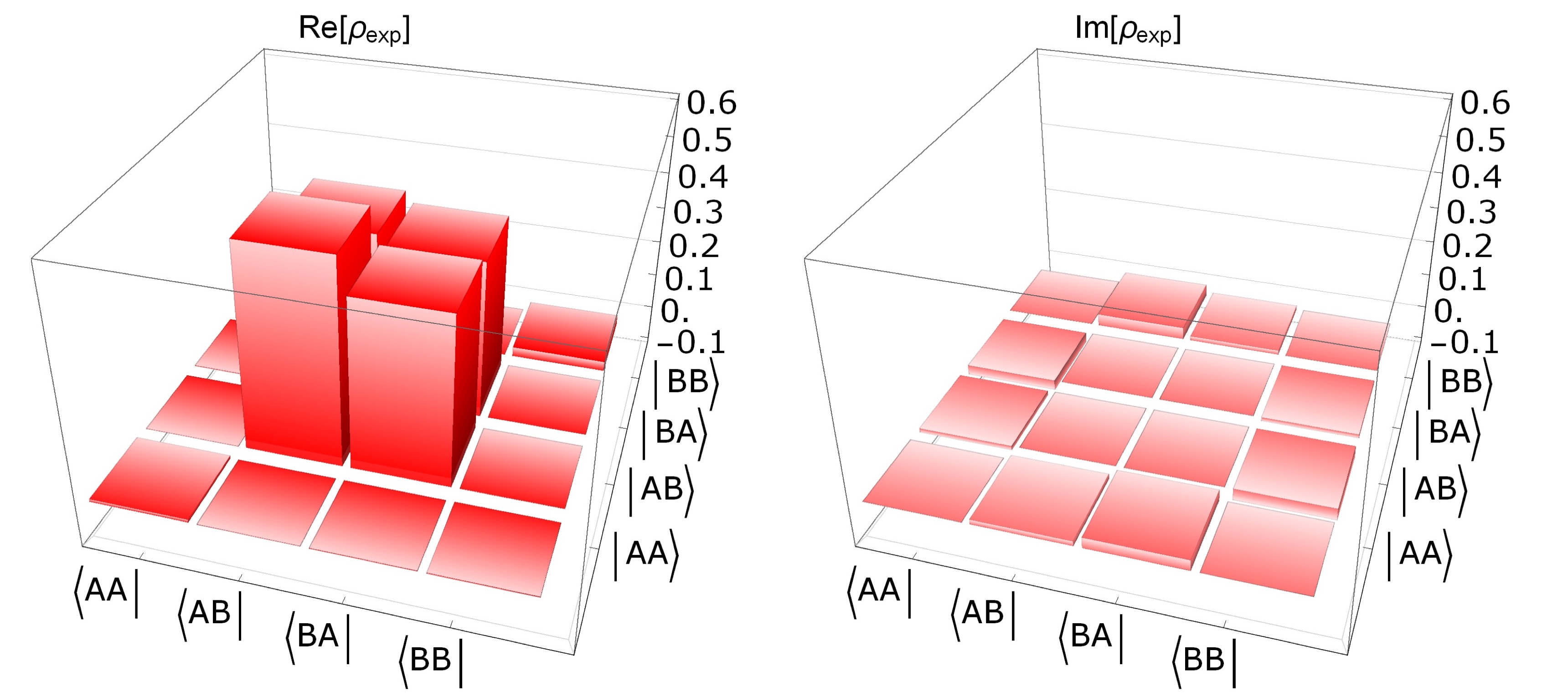} 	
	\caption{\emph{Reconstructed density matrix for path degree of freedom using SET: real part (left) and imaginary part (right).}}
 	\label{fig:pathSET}
 \end{figure}

In order to determine the density matrix by SET, we need to construct a proper seed beam that stimulates the pair generation. Given the particular geometry of our source, this can be done by removing the detector of the $\lambda_2$ photon and placing a cw laser operating at this wavelength, which inputs light in the very same single-mode fiber used to collect the photons in the spontaneous process (see Fig. \label{fig:source}). 
Parametric amplification at wavelength $\lambda_1$ is demonstrated by the enhancement of the photon generation rate by four orders of magnitude, with a count rate of almost $1$ MHz. A QWP, a HWP, a PS, and a BS are used to adjust the input parameters of the seed laser, as required by the SET protocol \cite{liscidini_13}.  The measurement time for the SET took 2-3 minutes in total. In particular, we modify the seed characteristics such that the light \emph{exiting} the setup mimics the properties (polarization and path) of the photon that would be detected in the corresponding QST measurements for SPDC.  This allows us to directly estimate the average number of pairs that would be detected in QST and calculate the corresponding density matrices.They are reported in Fig. \ref{fig:polarizationSET} and Fig. \ref{fig:pathSET} for polarization and path DOFs, respectively. Finally, fidelities, purities, tangles, and concurrences  are shown in Tab. \ref{tab:quantumparameters}.

The similarity between the density matrices obtained with SET and QST is such that a reader might be tempted to compare the results directly. Yet this should be done with care, for in general energy and momentum correlations with other DOFs are weighted differently in these two characterization approaches\cite{Rozema:15}. In SET,  pair emission is stimulated  using a cw laser, with the pairs being emitted and analyzed in a very narrow frequency range \textcolor{black}{(in our case the seed laser had a linewidth shorter than $0.001$nm)}. On the contrary, the smaller generation rates in QST typically require  the collection of the emitted pairs over much larger energy and momentum ranges. When this is properly taken into account, for example by performing energy-resolved SET measurements and averaging on the results, one finds complete agreement between the techniques \cite{Fang:16}. 

From these considerations, in our  experiment one expects some differences between SET and QST results for polarization, as we know from the QST characterization that correlations with other DOFs are present (see Eq. (\ref{pol_k})). Therefore, although at first glance the density matrices of Fig. \ref{fig:QST} and \ref{fig:SET} are very similar, it is not surprising that the purities, tangles, and concurrences obtained by SET differ from the values obtained using QST by 10\% to 25\%. The situation for the path DOF should be rather different, as from Eq. (\ref{hyper_k}) we do not expect significant correlations with other DOFs. And indeed, there is a high path purity shown with all the corresponding parameters for SET and QST in very good agreement, showing negligible discrepancies for tangle and concurrence, and differences of only 1-2\% for fidelity and purity. Such small differences may be attributed to systematic errors, which are not included in the uncertainties shown in Tab. \ref{tab:quantumparameters}. For example, an ideal implementation of SET requires all the seeded light to exit the source in the same mode \cite{liscidini_13}, while we verified that this is not so in our experiment, since some seed light is scattered by the apertures in the mask and by other optical elements of our setup.


In conclusion, we have demonstrated that SET can be used to characterize quantum states that are entangled in more than one degree of freedom, moving beyond start-of-the-art experiments that have only considered the polarization degree of freedom.  In particular, we have performed SET on a source generating photons hyperentangled in path and polarization, revealing all the main features of our complex system.  The demonstration of SET on the path DOF is a significant advance on the road to the implementation of this powerful technique on many different sources, including integrated devices, where DOFS different than polarization are used in quantum information studies and protocols, and where the SET advantages of speed and resolution will be most useful. 

\section{SUPPLEMENTAL:Purity and Hilbert spaces}
In this section we give more details about the connection between the purity of reduced density matrices associated with path and polarization degrees of freedom (DOFs). 

Since the measured purity of the reduced density operator characterizing the path degree of freedom is close to unity, it is reasonable to consider the full state of the system as essentially pure. Then the simplest assumption would be that one could consider the relevant Hilbert space to be a direct product of a Hilbert space associated with the path of the photons and one associated with their polarization, 
\begin{equation}
\mathcal{H}=\mathcal{H}_\mathrm{Pol}\otimes\mathcal{H}_\mathrm{Path},
\end{equation}
but this will not suffice. Indeed, a ket in this Hilbert space could always be Schmidt decomposed, 
\begin{equation}
\left|\psi\right\rangle =\sum_{i}\alpha_{i}\left|Path_{i}\right\rangle \left|Pol_{i}\right\rangle ,
\end{equation}
with
\begin{eqnarray}
\left\langle Path_{i}|Path_{j}\right\rangle =\delta_{ij},\\\left\langle Pol_{i}|Pol_{j}\right\rangle =\delta_{ij},\\\sum_{i}\left|\alpha_{i}\right|^{2}=1,
\end{eqnarray}
and we would have
\begin{eqnarray}
&\mathrm{Tr_{Pol}}\left[\rho_{\mathrm{Pol}}^{2}\right]=\mathrm{Tr_{Pol}}\left[\left(\mathrm{Tr_{Path}}\left[|\psi\rangle\langle\psi|\right]\right)^{2}\right]=\sum_{i}\left|\alpha_{i}\right|^{4},\\&\mathrm{Tr_{Path}}\left[\rho_{\mathrm{Path}}^{2}\right]=\mathrm{Tr_{Path}}\left[\left(\mathrm{Tr_{Pol}}\left[|\psi\rangle\langle\psi|\right]\right)^{2}\right]=\sum_{i}\left|\alpha_{i}\right|^{4},
\end{eqnarray}
while experimentally the purity of the reduced density operators associated with path and polarization are significantly different. 

Thus we consider the experimental scenario in more detail, and take into account the finite size of the holes in the mask. In the simple argument above, these holes were associated with the Hilbert space of the \emph{path} of the photons. But in fact, photons with different momenta will be collected by the same hole, and the mask will then effectively lead to a ?trace? over those momenta, with part of information related to the emission direction lost once the photons are collected by the fiber. In particular, this information is related to small deviations from the propagation direction given by the center of the nonlinear crystal and that of the holes in the mask. For these reasons, we start with the generated state 
\begin{equation}
|\psi\rangle=\sum_{l,m}\int d\mathbf{k_{i}}d\mathbf{\mathbf{k_{s}}}\left[\phi_{l,m}^{fwd}(\mathbf{k_{i}},\mathbf{k_{s}})+\phi_{l,m}^{\mathrm{ref}}(\mathbf{k_{i}},\mathbf{k_{s}})\right]\left|\mathbf{k_{i}},l\right\rangle \left|\mathbf{k_{s}},m\right\rangle ,
\end{equation}
where $\phi_{l,m}^{fwd(ref)}(\mathbf{k_{i}},\mathbf{k_{s}})$ is the biphoton wave function associated with the pair generated for a forward (reflected) pump, where $l,m\in(V,H)$, and $\left|\mathbf{k_{i}},l\right\rangle$  is the idler photon state of momentum $\mathbf{k_{i}}$ and polarization $m$, etc. 

The mask is described by the projector
\begin{eqnarray}\nonumber
&P_{mask}&\\ \nonumber
&=&\sum_{\mathbf{K_{i}},\mathbf{K_{s}}}\int_{\Omega}d\kappa_{i}d\kappa_{s}\left|\mathbf{K_{i}}+\kappa_{i}\right\rangle  \left|\mathbf{K_{s}}+\kappa_{s}\right\rangle \left\langle \mathbf{K_{s}}+\kappa_{s}\right|\left\langle \mathbf{K_{i}}+\kappa_{i}\right|,\\
\end{eqnarray}
where $\mathbf{K_{i(s)}}$ corresponds to the positions $T$ on the mask, left (L) or right (R) with respect to the center of the holes for idler and signal (see Fig. 1 (a); $\mathbf{K}_{i}=\mathbf{K}_{i}(R)$ or $\mathbf{K}_{i}(L)$, for example), while $\Omega$ indicates the 2-dimensional integration range over $\kappa_{i(s)}$ determined by the size of the holes in the mask, $\kappa_{i,s}=0$ corresponds to the hole centers. This, leads to the state:
\begin{eqnarray}\nonumber
|\psi_{mask}\rangle&=&\sum_{l,m}\sum_{\mathbf{K_{i}},\mathbf{K_{s}}}\int_{\Omega}d\kappa_{i}d\kappa_{s}\\ \nonumber
&\times&\left[\phi_{l,m}^{fwd}(\mathbf{K_{i}}+\kappa_{i},\mathbf{K_{s}}+\kappa_{s})+\phi_{l,m}^{\mathrm{ref}}(\mathbf{K_{i}}+\kappa_{i},\mathbf{K_{s}}+\kappa_{s})\right]\\
&\times&\left|\mathbf{K_{i}}+\kappa_{i},l\right\rangle \left|\mathbf{K_{s}}+\kappa_{s},m\right\rangle.
\end{eqnarray}
Here the Hilbert space can be considered a direct product of Hilbert spaces associated with the paths, the polarization, and the distribution of momenta within each hole that identifies the paths, 
\begin{equation}
\mathcal{H} =\mathcal{H}_\mathrm{Pol}\otimes\mathcal{H}_\mathrm{\mathrm{Path}}\otimes\mathcal{H}_{\kappa},
\end{equation}
where $\mathcal{H_{\mathrm{Pol}}}$ is the polarization Hilbert space, $\mathcal{H}_{\mathrm{\mathrm{Path}}}$ the path Hilbert space, which depends on the hole position through $\mathbf{K}_{i}$ and $\mathbf{K}_{s}$, and $\mathcal{H}_\kappa$ the Hilbert space associated with the momentum DOF determined by the hole size. This factorization is natural  because one can define the operators such as $a_{i,l,T,\kappa}^{\dagger}$, which describes the creation of an idler photon exiting in the path $T$, having polarization $l$,  and with momentum in the direction given by $\mathbf{K}_i(T)+\kappa$. These operators satisfy the commutation relations
\begin{eqnarray}
\left[a_{i,(l,T,\kappa)},a_{i,(l',T,\kappa')}^{\dagger}\right]&=&\delta_{l,l'}\delta_{T,T'}\delta(\kappa-\kappa'),
\\
\left[a_{s(l,T,\kappa)},a_{s,(l',T,\kappa')}^{\dagger}\right]&=&\delta_{l,l'}\delta_{T,T'}\delta(\kappa-\kappa'),
\\
\left[a_{i,(l,T,\kappa)},a_{s,(l',T,\kappa')}^{\dagger}\right]&=&0.
\end{eqnarray}
Thus, we can rewrite
\begin{eqnarray}\nonumber
|\psi_{mask}\rangle&=&\sum_{l,l'}\sum_{T,T'}\int_{\Omega}d\kappa d\kappa'\left[\phi_{l,l',T,T'}^{fwd}(\kappa,\kappa')+\phi_{l,l',T,T'}^{\mathrm{ref}}(\kappa,\kappa')\right]\\ &\times&\left|T,\kappa,l\right\rangle _{i}\left|T',\kappa',l'\right\rangle _{s}.
\end{eqnarray}
where $\left|(T,\kappa,l)i;(T',\kappa',l')s\right\rangle =a_{(T,l,\kappa)i}^{\dagger}a_{(T',l',\kappa')s}^{\dagger}\left|vac\right\rangle$. Given the symmetry of our source and mask, we can safely assume that polarization and $\kappa$ DOFs are independent of the path, and take 
\begin{equation}
\phi_{l,l',T,T'}^{\mathrm{ref(fwd)}}(\kappa,\kappa')=(\delta_{T,A}\delta_{T'B}+e^{\imath\theta}\delta_{T,B}\delta_{T',A})f_{l,l'}^{\mathrm{ref(fwd)}}(\kappa,\kappa'),
\end{equation}
which allows us to write
\begin{equation}
|\psi_{mask}\rangle=|\psi_\mathrm{path}\rangle\otimes|\psi_{\mathrm{pol},\kappa}\rangle,
\end{equation}
with
\begin{equation}
\label{eq:pathstate}
\ket{\psi_{\mathrm{path}}}=\frac{1}{\sqrt{2}}(\ket{A_{\lambda_{1}}B_{\lambda_{2}}}+\text{e}^{i\theta}\ket{B_{\lambda_{1}}A_{\lambda_{2}}}),
\end{equation}
where the $L$ and $R$ states are denoted by $A$ and $B$, and 
\begin{equation}
|\psi_{\mathrm{pol},\kappa}\rangle=\sum_{l,l'}\int_{\Omega}d\kappa d\kappa'\left[f_{l,l'}^{fwd}(\kappa,\kappa')+f_{l,l'}^{\mathrm{ref}}(\kappa,\kappa')\right] \left|\kappa,l\right\rangle _{i}\left|\kappa',l'\right\rangle _{s}.
\end{equation}
We notice that, while the purity of the reduced path density operator  given by \eqref{eq:pathstate} is unity, in general the purity of the polarization reduced density operator is less than unity, because of the correlations between the polarization and the collection angle associated with the size of the holes in the mask.

\section{Visibility and Purity}
In the following section we justify how the expected purity for a state encoded in the path degree of freedom can be lower than unity, if we accounted for imperfections of the experimental apparatus. 
We describe the link between the visibility of the interference of a pure quantum state entering a lossless unbalanced Beam Splitter (LUBS) and the purity of a mixed state that enters in an lossless balanced Beam Splitter (LBBS) and generates the same visibility. In this way we can calculate the maximum expected value of the purity for a state entering a lossless unbalanced BS.
\subsection{1 qubit}
\subsubsection{Visibility with a pure state and unbalanced beam splitter}
\begin{figure}[h]
	\centering
	\includegraphics[scale=0.25]{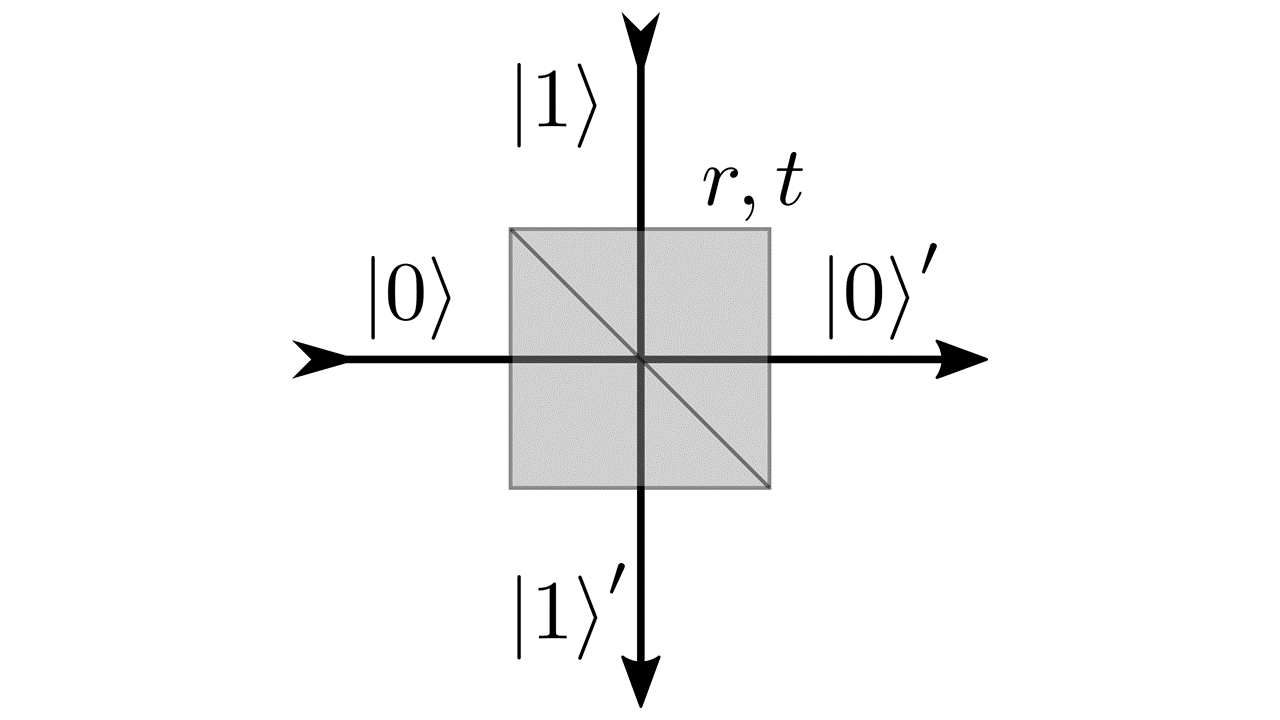}
	\caption{\emph{Sketch of LUBS and eigenstates of the path degree of freedom.}}
	\label{fig:BS}
\end{figure}

Suppose we have a qubit encoded in path degree of freedom and that the state of the qubit is represented by the normalized ket
\begin{equation}
	\ket{\psi}=\frac{1}{\sqrt{2}}(\ket{0}+\text{e}^{i\phi}\ket{1})
\end{equation}
where $\ket{0}$ and $\ket{1}$ are the eigenstates of the path and the input arms of a LUBS (see Fig.\ref{fig:BS}).


The LUBS is characterized by a transmittivity coefficient $t$ and a reflectivity coefficient $r$ ($t,r \in \mathbb{R}^+$), such that
\begin{equation}
\label{eq:lossless}
	t^2+r^2=1.
\end{equation}
The LUBS acts on $\ket{0}$, $\ket{1}$ as follow:
\begin{equation}
	\begin{split}
		&\hat{BS}\ket{0}=t\ket{0}'+ir\ket{1}'\\
		&\hat{BS}\ket{1}=t\ket{1}'+ir\ket{0}'
	\end{split}
\end{equation}
where the kets $\ket{0}'$ and $\ket{1}'$ represent the output modes of the LUBS. The evolution of $\ket{\psi}$ is:
\begin{equation}
	\ket{\psi}'=\frac{1}{\sqrt{2}}[\ket{0}'(t+i\text{e}^{i\phi}r)+\ket{1}'(ir+\text{e}^{i\phi}t)].
\end{equation}
The probability of obtaining $\ket{0}'$ after a measurement, for instance, depends on the phase $\phi$,  and it is maximized (minimized), when $\phi=-\frac{\pi}{2}$ ($\phi=\frac{\pi}{2}$), such that: 
\begin{equation}
	\begin{split}
		&Max=M=\frac{(t+r)^2}{2}\\
		&min=m=\frac{(t-r)^2}{2}.
	\end{split}
\end{equation} 
The visibility V of the interference in such case is:
\begin{equation}
\label{eq:vlubs}
	V_{LUBS}=\frac{M-m}{M+m}=\frac{t^2+r^2+2rt-t^2-r^2+2rt}{t^2+r^2+2rt+t^2+r^2-2rt}=\frac{4rt}{2}=2rt
\end{equation} 
where we have used the relation (\ref{eq:lossless}) in the third passage. If $t\ne r\ne \frac{1}{\sqrt{2}}$, then $V\ne 1$. This result shows the dependence of the visibility to the parameters of the LUBS. 
\subsubsection{Purity of a mixed quantum state in a ideal beam splitter }
We now consider a mixed state as input of an LBBS, 
namely:
\begin{equation}
\label{eq:rho(p)}
\hat{\rho}=(1-p)\ket{\psi}\bra{\psi}+\frac{p}{2}\mathcal{I},
\end{equation}
where $\mathcal{I}$ is the 2 dimensional identity matrix and p is a parameter connected to the purity of the state. The state after an LBBS ($t=r=\frac{1}{\sqrt{2}}$) becomes:
\begin{equation}
	\begin{split}
		\hat{BS}\hat{\rho}\hat{BS}^\dagger=&
		\frac{1-p}{4}
		\begin{pmatrix}
		|1+i\text{e}^{i\phi}|^2 & (1+i\text{e}^{i\phi})(\text{e}^{-i\phi}-i)\\
		(1-i\text{e}^{-i\phi})(\text{e}^{i\phi}+i) & |\text{e}^{i\phi}+i|^2	
		\end{pmatrix}\\+ 
		&\frac{p}{2}
		\begin{pmatrix}
			1 & 0\\
			0 & 1
		\end{pmatrix}.
	\end{split}
\end{equation}
We proceed as before, evaluating the probability of obtaining state $\ket{0'}$ after a measurement of the system, and then we calculate the visibility $V_{\hat{\rho}}$, such that:
\begin{equation}
	V_{\hat{\rho}}=\frac{max-min}{max+min}=\frac{1-p+\frac{p}{2}-\frac{p}{2}}{1-p+\frac{p}{2}+\frac{p}{2}}=1-p.
\end{equation}
We can then substitute the value of $p$ in Eq. \ref{eq:rho(p)} and obtain
\begin{equation}
	\begin{split}
	\hat{\rho}&=V_{\hat{\rho}}\frac{(\ket{0}+\text{e}^{i\phi}\ket{1})(\bra{0}+\text{e}^{-i\phi}\bra{1})}{2}+(1-V_{\hat{\rho}})\frac{\ket{0}\bra{0}+\ket{1}\bra{1}}{2}\\
    &=\frac{V_{\hat{\rho}}}{2}
	\begin{pmatrix}
	1 & \text{e}^{-i\phi}\\
	\text{e}^{i\phi} & 1
	\end{pmatrix}
	+ \frac{1-V_{\hat{\rho}}}{2}
	\begin{pmatrix}
	1 & 0\\
	0 & 1
	\end{pmatrix} = \begin{pmatrix}
	\frac{1}{2} & \text{e}^{-i\phi}\frac{V}{2}\\
	\text{e}^{i\phi}\frac{V}{2} & \frac{1}{2}
	\end{pmatrix}.
	\end{split}
\end{equation}
From this, we recover the purity P of the state, as a function of $V_{\hat{\rho}}$:
\begin{equation}
\label{eq:linkVP}
	\begin{split}
	\hat{\rho}^2&=\begin{pmatrix}
	\frac{1}{2} & \text{e}^{-i\phi}\frac{V}{2}\\
	\text{e}^{i\phi}\frac{V}{2} & \frac{1}{2}
	\end{pmatrix} \begin{pmatrix}
	\frac{1}{2} & \text{e}^{-i\phi}\frac{V}{2}\\
	\text{e}^{i\phi}\frac{V}{2} & \frac{1}{2}
	\end{pmatrix}\\
    &=\begin{pmatrix}
	\frac{1}{4}+\frac{V^2}{4} & \text{e}^{-i\phi}\frac{V}{2}\\
	\text{e}^{i\phi}\frac{V}{2} & \frac{1}{4}+\frac{V^2}{4}
	\end{pmatrix}
	\end{split}
\end{equation}
and
\begin{equation}
\label{eq:purity}
	P=Tr(\hat{\rho}^2)=\frac{1}{2}+\frac{V^2}{2}.
\end{equation}
\subsubsection{Bounds on the purity of a state, given an experimental visibility}
The previous results show that can have a state with $P\leq 1$ if we have a LUBS. Indeed, combining \ref{eq:purity} and \ref{eq:vlubs}, and assuming $V_{\hat{\rho}}=V_{LUBS}$, we find that $P_{exp}\leq \frac{1}{2}+\frac{V^2}{2}$.
We can say that if $V\ne 1$ the state that is measured by our non-ideal setup is a mixed state with maximum purity given by Eq. (\ref{eq:linkVP}).

\subsection{2 qubits}
\begin{figure}[h]
	\centering
	\includegraphics[scale=0.25]{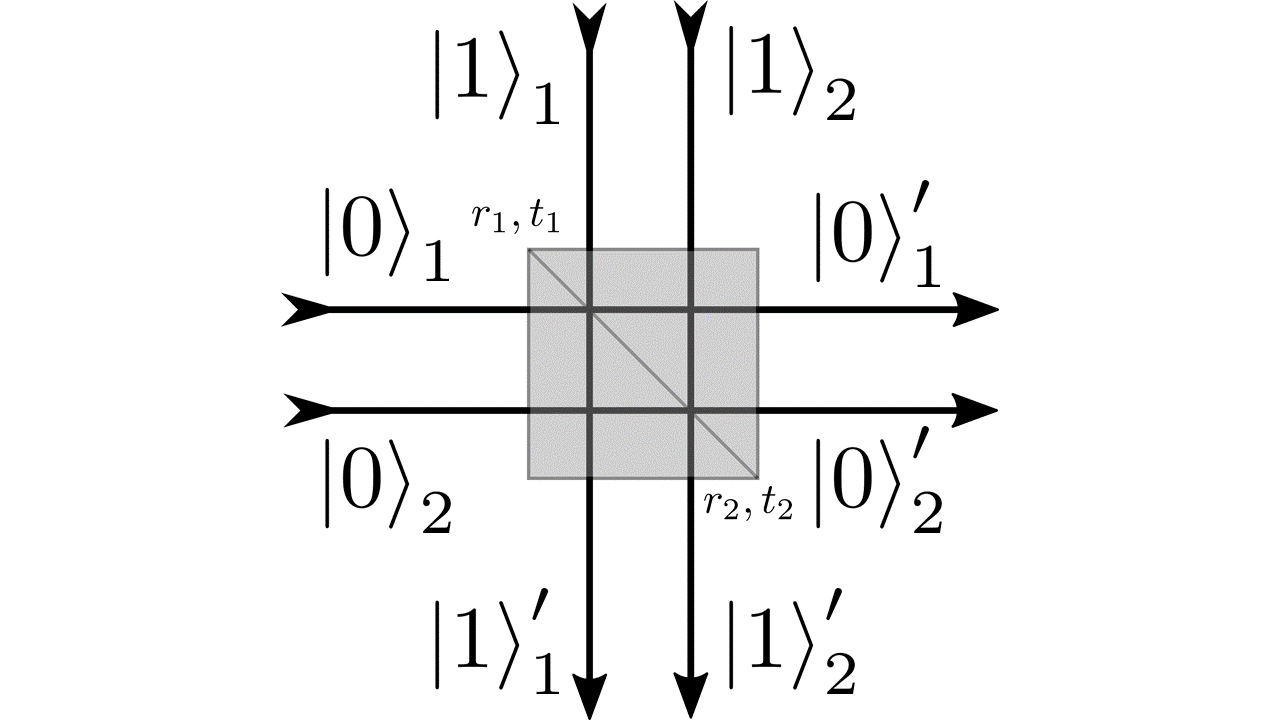}
	\caption{\emph{Sketch of the BS and the eigenstates of the path degree of freedom for both qubit 1 and 2.}}
	\label{fig:BS2qubit}
\end{figure}

Now we address the 2-qubit case. Suppose we have a path entangled state
\begin{equation}
	\ket{\psi}=\frac{1}{\sqrt{2}}(\ket{0}_1\ket{1}_2+\text{e}^{i\phi}\ket{1}_1\ket{0_2})=\frac{1}{\sqrt{2}}(\ket{01}+\text{e}^{i\phi}\ket{10})
\end{equation}
where the label 1 (2) represent the first (second) qubit. We consider the eigenstates $\ket{0}, \ket{1}$ as kets representing the input modes of two LUBSs (see Fig.\ref{fig:BS2qubit}).
\\Each LUBS has its own values of $r$ and $t$:
\begin{equation}
	\begin{split}
		&\hat{BS}_1\ket{0}=t_1\ket{0}'+ir_1\ket{1}'\\
		&\hat{BS}_1\ket{1}=t_1\ket{1}'+ir_1\ket{0}'\\
		&\hat{BS}_2\ket{0}=t_2\ket{0}'+ir_2\ket{1}'\\
		&\hat{BS}_2\ket{1}=t_2\ket{1}'+ir_2\ket{0}'	
	\end{split}
\end{equation}
where $r_1,r_2,r_1,t_2 \in \mathbb{R}^+$.\\
After the interaction with the LUBS we obtain
\begin{equation}
	\begin{split}
		\ket{\psi}'&=\frac{1}{\sqrt{2}}[(t_1\ket{0}'+ir_1\ket{1}')(t_2\ket{1}'+ir_2\ket{0}')\\
        &+\text{e}^{i\phi}(t_1\ket{1}'+ir_1\ket{0}')(t_2\ket{0}'+ir_2\ket{1}')]\\
        &=\frac{1}{\sqrt{2}}(t_1t_2\ket{01}'+it_1r_2\ket{00}'+ir_1t_2\ket{11}'-r_1r_2\ket{10}'\\
        &+\text{e}^{i\phi}t_1t_2\ket{10}'
  +i\text{e}^{i\phi}t_1r_2\ket{11}'+i\text{e}^{i\phi}r_1t_2\ket{00}'-\text{e}^{i\phi}r_1r_2\ket{01}')\\
  &=\frac{1}{\sqrt{2}}[\ket{01}'(t_1t_2-\text{e}^{i\phi}r_1r_2)+i\ket{00}'(t_1r_2+\text{e}^{i\phi}r_1t_2)\\
        &+i\ket{11}'(r_1t_2+\text{e}^{i\phi}t_1r_2)-\ket{10}'(r_1r_2-\text{e}^{i\phi}t_1t_2)]
	\end{split}.
\end{equation}
If we look at one of the four elements of the state, for example at the ket $\ket{01}'$, we can evaluate the maximum and minimum value of the interference
\begin{equation}
	\begin{split}
		&Max=M=\frac{(t_1t_2+r_1r_2)^2}{2}\\
		&min=m=\frac{(t_1t_2-r_1r_2)^2}{2}
	\end{split}
\end{equation}
and the visibility
\begin{equation}
	\begin{split}
	V&=\frac{M-m}{M+m}=\frac{(t_1t_2+r_1r_2)-(t_1t_2-r_1r_2)}{(t_1t_2+r_1r_2)+(t_1t_2-r_1r_2)}\\
    &=1-\frac{(t_1t_2-r_1r_2)^2}{t_1^2t_2^2+r_1^2r_2^2}.
	\end{split}
\end{equation}\\
\underline{\hspace{8.8cm}}\\
This show that one can obtain visibility different from unity even if the input state is pure state. For this reason we want to find a mixed state that gives the same visibility when interacting with LBBS and then find a link between the visibility and the purity of the mixed state.\\
In order to do that we write the state as
\begin{equation}
\label{eq:rho(p)2qubits}
	\begin{split}
		\hat{\rho}&=(1-p)\frac{(\ket{01}+\text{e}^{i\phi}\ket{10})(\bra{01}+\text{e}^{-i\phi}\bra{10})}{2}\\
        &+p\frac{\ket{01}\bra{01}+\ket{10}\bra{10}}{2}.
	\end{split}
\end{equation}
In order to determine the value of p we apply the LBBS transformation to the state $\rho$
\begin{widetext}
\begin{equation}
	\begin{split}
	&(\hat{BS}\otimes\hat{BS})\hat{\rho}(\hat{BS}^{\dagger}\otimes\hat{BS}^{\dagger})=\\
		&\frac{1-p}{8}
		\begin{pmatrix}
		|1+\text{e}^{i\phi}|^2 & i(1+\text{e}^{i\phi})(1-\text{e}^{-i\phi}) & -i(1+\text{e}^{i\phi})(1-\text{e}^{-i\phi}) & |1+\text{e}^{i\phi}|^2\\
		-i(1+\text{e}^{-i\phi})(1-\text{e}^{i\phi}) & |1-\text{e}^{i\phi}|^2 & -|1-\text{e}^{i\phi}|^2 & -i(1-\text{e}^{i\phi})(1+\text{e}^{-i\phi})\\
		i(1-\text{e}^{i\phi})(1+\text{e}^{-i\phi}) & -|1-\text{e}^{i\phi}|^2 & |1-\text{e}^{i\phi}|^2 & i(1-\text{e}^{i\phi})(1+\text{e}^{-i\phi})\\
		|1+\text{e}^{i\phi}|^2 & i(1+\text{e}^{i\phi})(1-\text{e}^{-i\phi}) & -i(1+\text{e}^{i\phi})(1-\text{e}^{-i\phi}) & |1+\text{e}^{i\phi}|^2
		\end{pmatrix} +
		\frac{p}{4}
		\begin{pmatrix}
		1 & 0 & 0 & 1 \\
		0 & 1 & -1 & 0 \\
		0 & -1 & 1 & 0 \\
		1 & 0 & 0 & 1 
		\end{pmatrix}.		
	\end{split}
\end{equation}
\end{widetext}
Now we impose that the visibility of interference of $\hat{\rho}$ on the LBBS is equal to the one of the pure state on the LUBS. The maximum reached after the LBBS is the sum of the maximum of the interference due to the pure state contained in $\hat{\rho}$ and the contribution of the mixed state of $\hat{\rho}$, while the minimum is given only by the mixed state because for the pure state the interference is complete. So
\begin{equation}
	\begin{split}
	\frac{maximum-minimum}{maximum+minimum}&=\frac{\frac{1-p}{2}+\frac{p}{4}-\frac{p}{4}}{\frac{1-p}{2}+\frac{p}{4}+\frac{p}{4}}=\frac{\frac{1-p}{2}}{\frac{1}{2}}=1-p\overset{!}{=}V
    \\&\longrightarrow p=1-V.
	\end{split}
\end{equation}
We can then substitute the value of $p$ in Eq. (\ref{eq:rho(p)2qubits}) and obtain
\begin{equation}
	\begin{split}
		\hat{\rho}&=V\frac{(\ket{01}+\text{e}^{i\phi}\ket{10})(\bra{01}+\text{e}^{-i\phi}\bra{10})}{2}\\
        &+(1-V)\frac{\ket{01}\bra{01}+\ket{10}\bra{10}}{2}\\
        &=\quad\frac{V}{2}\begin{pmatrix}
		0 & 0 & 0 & 0\\
		0 & 1 & \text{e}^{-i\phi} & 0\\
		0 & \text{e}^{i\phi} & 1 & 0\\
		0 & 0 & 0 & 0
		\end{pmatrix} + 
		\frac{1-V}{2}
		\begin{pmatrix}
		0 & 0 & 0 & 0\\
		0 & 1 & 0 & 0\\
		0 & 0 & 1 & 0\\
		0 & 0 & 0 & 0
		\end{pmatrix}\\
        &=
		\begin{pmatrix}
		0 & 0 & 0 & 0\\
		0 & \frac{1}{2} & \text{e}^{-i\phi}\frac{V}{2} & 0\\
		0 & \text{e}^{i\phi}\frac{V}{2} & \frac{1}{2} & 0\\
		0 & 0 & 0 & 0
		\end{pmatrix}.
	\end{split}
\end{equation}\\
\\
\underline{\hspace{8.8cm}}\\
Now we can calculate the square of the density matrix
\begin{equation}
	\begin{split}
	\hat{\rho}^2&=\begin{pmatrix}
	0 & 0 & 0 & 0\\
	0 & \frac{1}{2} & \text{e}^{-i\phi}\frac{V}{2} & 0\\
	0 & \text{e}^{i\phi}\frac{V}{2} & \frac{1}{2} & 0\\
	0 & 0 & 0 & 0
	\end{pmatrix}
	\begin{pmatrix}
	0 & 0 & 0 & 0\\
	0 & \frac{1}{2} & \text{e}^{-i\phi}\frac{V}{2} & 0\\
	0 & \text{e}^{i\phi}\frac{V}{2} & \frac{1}{2} & 0\\
	0 & 0 & 0 & 0
	\end{pmatrix}\\
    &=
	\begin{pmatrix}
	0 & 0 & 0 & 0\\
	0 & \frac{1}{4}+\frac{V^2}{4} & \text{e}^{-i\phi}\frac{V^2}{2} & 0\\
	0 & \text{e}^{i\phi}\frac{V^2}{2} & \frac{1}{4}+\frac{V^2}{4} & 0\\
	0 & 0 & 0 & 0
	\end{pmatrix}
	\end{split}
\end{equation}
and the purity
\begin{equation}
\label{eq:linkVP2qubits}
	Tr(\hat{\rho}^2)=\frac{1}{2}+\frac{V^2}{2}.
\end{equation}
So the link between the visibility given by a pure state entering an unbalanced LUBS and the purity of a mixed state that enters an LBBS and that gives the same visibility is given by the Eq. (\ref{eq:linkVP2qubits}). In this sense we can say that if $V\ne 1$ the state that is measured by our imperfect setup is a mixed state with maximum purity given by Eq. (\ref{eq:linkVP2qubits}).

We can then evaluate the maximum purity that can be reached with the BS used in our experiment. In our case the subscript $1,2$ of $r_{1,2}$ and $t_{1,2}$ denotes the different wavelengths of the photons. The numerical values of the BS's parameters are listed in Table \ref{tab:BSparameters}.
\begin{table}[!h]
	\centering
	\begin{tabular}{ccc} {\em Parameter}&{\em $633$nm}&{\em $770$nm}\\ \hline
		$r_H$ & $\sqrt{0.42}$ & $\sqrt{0.45}$ \\ \hline
		$t_H$ & $\sqrt{0.58}$ & $\sqrt{0.55}$ \\ \hline
		$r_V$ & $\sqrt{0.36}$ & $\sqrt{0.43}$ \\ \hline
		$t_V$ & $\sqrt{0.64}$ & $\sqrt{0.57}$ \\ \hline
	\end{tabular}	
	\caption{\emph{Parameters of the BS used in the SET experiment.}}
	\label{tab:BSparameters}
\end{table}
We obtain, through the Eq.(\ref{eq:linkVP2qubits}), the maximum purity for horizontally and vertically polarized photons
\begin{equation}
	\begin{split}
		&Tr(\hat{\rho}^2)_{H}=96.725\%\\
		&Tr(\hat{\rho}^2)_{V}=91.833\%
	\end{split}
\end{equation}
The values of the purity measured experimentally are
\begin{equation}
	\begin{split}
		&Tr(\hat{\rho}^2)_{QST}=(90.9\pm0.3)\%\\
		&Tr(\hat{\rho}^2)_{SET}=(88.6\pm0.1)\%
	\end{split}
\end{equation}
The experimental data are compatible with these theoretical predictions, which are upper bounds for the purity. 

\section{Measurement time}
As mentioned in the manuscript, to the best of our knowledge, SET in a DOF other than polarization has never been attempted before. Thus we decided to use a source that could be characterized via traditional QST.  As in the case of Ref.11, in which SET has been performed on a bulk crystal, here there is not significant advantage in terms of the time measurement, nor resolution. However, this choice gave the reader the opportunity to compare the results of SET and QST. 
The measurement time for the SET in both DOFs, polarization and path,  took 2-3 minutes in total, while QST required about 10 minutes.


\begin{thebibliography}{23}%
\makeatletter
\providecommand \@ifxundefined [1]{%
 \@ifx{#1\undefined}
}%
\providecommand \@ifnum [1]{%
 \ifnum #1\expandafter \@firstoftwo
 \else \expandafter \@secondoftwo
 \fi
}%
\providecommand \@ifx [1]{%
 \ifx #1\expandafter \@firstoftwo
 \else \expandafter \@secondoftwo
 \fi
}%
\providecommand \natexlab [1]{#1}%
\providecommand \enquote  [1]{``#1''}%
\providecommand \bibnamefont  [1]{#1}%
\providecommand \bibfnamefont [1]{#1}%
\providecommand \citenamefont [1]{#1}%
\providecommand \href@noop [0]{\@secondoftwo}%
\providecommand \href [0]{\begingroup \@sanitize@url \@href}%
\providecommand \@href[1]{\@@startlink{#1}\@@href}%
\providecommand \@@href[1]{\endgroup#1\@@endlink}%
\providecommand \@sanitize@url [0]{\catcode `\\12\catcode `\$12\catcode
  `\&12\catcode `\#12\catcode `\^12\catcode `\_12\catcode `\%12\relax}%
\providecommand \@@startlink[1]{}%
\providecommand \@@endlink[0]{}%
\providecommand \url  [0]{\begingroup\@sanitize@url \@url }%
\providecommand \@url [1]{\endgroup\@href {#1}{\urlprefix }}%
\providecommand \urlprefix  [0]{URL }%
\providecommand \Eprint [0]{\href }%
\providecommand \doibase [0]{http://dx.doi.org/}%
\providecommand \selectlanguage [0]{\@gobble}%
\providecommand \bibinfo  [0]{\@secondoftwo}%
\providecommand \bibfield  [0]{\@secondoftwo}%
\providecommand \translation [1]{[#1]}%
\providecommand \BibitemOpen [0]{}%
\providecommand \bibitemStop [0]{}%
\providecommand \bibitemNoStop [0]{.\EOS\space}%
\providecommand \EOS [0]{\spacefactor3000\relax}%
\providecommand \BibitemShut  [1]{\csname bibitem#1\endcsname}%
\let\auto@bib@innerbib\@empty
\bibitem [{\citenamefont {Altepeter}\ \emph
  {et~al.}(2005{\natexlab{a}})\citenamefont {Altepeter}, \citenamefont
  {Jeffrey},\ and\ \citenamefont {Kwiat}}]{kwiat:tomography}%
  \BibitemOpen
  \bibfield  {author} {\bibinfo {author} {\bibfnamefont {J.~B.}\ \bibnamefont
  {Altepeter}}, \bibinfo {author} {\bibfnamefont {E.~R.}\ \bibnamefont
  {Jeffrey}}, \ and\ \bibinfo {author} {\bibfnamefont {P.~G.}\ \bibnamefont
  {Kwiat}},\ }\href@noop {} {\bibfield  {journal} {\bibinfo  {journal}
  {Advances in Atomic, Molecular, and Optical Physics}\ }\textbf {\bibinfo
  {volume} {52}} (\bibinfo {year} {2005}{\natexlab{a}})}\BibitemShut {NoStop}%
\bibitem [{\citenamefont {White}\ \emph {et~al.}(1999)\citenamefont {White},
  \citenamefont {James}, \citenamefont {Eberhard},\ and\ \citenamefont
  {Kwiat}}]{white_99}%
  \BibitemOpen
  \bibfield  {author} {\bibinfo {author} {\bibfnamefont {A.~G.}\ \bibnamefont
  {White}}, \bibinfo {author} {\bibfnamefont {D.~F.~V.}\ \bibnamefont {James}},
  \bibinfo {author} {\bibfnamefont {P.~H.}\ \bibnamefont {Eberhard}}, \ and\
  \bibinfo {author} {\bibfnamefont {P.~G.}\ \bibnamefont {Kwiat}},\ }\href
  {\doibase 10.1103/PhysRevLett.83.3103} {\bibfield  {journal} {\bibinfo
  {journal} {Phys. Rev. Lett.}\ }\textbf {\bibinfo {volume} {83}},\ \bibinfo
  {pages} {3103} (\bibinfo {year} {1999})}\BibitemShut {NoStop}%
\bibitem [{\citenamefont {Barreiro}\ \emph {et~al.}(2005)\citenamefont
  {Barreiro}, \citenamefont {Langford}, \citenamefont {Peters},\ and\
  \citenamefont {Kwiat}}]{Barreiro_05}%
  \BibitemOpen
  \bibfield  {author} {\bibinfo {author} {\bibfnamefont {J.~T.}\ \bibnamefont
  {Barreiro}}, \bibinfo {author} {\bibfnamefont {N.~K.}\ \bibnamefont
  {Langford}}, \bibinfo {author} {\bibfnamefont {N.~A.}\ \bibnamefont
  {Peters}}, \ and\ \bibinfo {author} {\bibfnamefont {P.~G.}\ \bibnamefont
  {Kwiat}},\ }\href {\doibase 10.1103/PhysRevLett.95.260501} {\bibfield
  {journal} {\bibinfo  {journal} {Phys. Rev. Lett.}\ }\textbf {\bibinfo
  {volume} {95}},\ \bibinfo {pages} {260501} (\bibinfo {year}
  {2005})}\BibitemShut {NoStop}%
\bibitem [{\citenamefont {Boitier}\ \emph {et~al.}(2014)\citenamefont
  {Boitier}, \citenamefont {Orieux}, \citenamefont {Autebert}, \citenamefont
  {Lema\^{\i}tre}, \citenamefont {Galopin}, \citenamefont {Manquest},
  \citenamefont {Sirtori}, \citenamefont {Favero}, \citenamefont {Leo},\ and\
  \citenamefont {Ducci}}]{Ducci_14}%
  \BibitemOpen
  \bibfield  {author} {\bibinfo {author} {\bibfnamefont {F.}~\bibnamefont
  {Boitier}}, \bibinfo {author} {\bibfnamefont {A.}~\bibnamefont {Orieux}},
  \bibinfo {author} {\bibfnamefont {C.}~\bibnamefont {Autebert}}, \bibinfo
  {author} {\bibfnamefont {A.}~\bibnamefont {Lema\^{\i}tre}}, \bibinfo {author}
  {\bibfnamefont {E.}~\bibnamefont {Galopin}}, \bibinfo {author} {\bibfnamefont
  {C.}~\bibnamefont {Manquest}}, \bibinfo {author} {\bibfnamefont
  {C.}~\bibnamefont {Sirtori}}, \bibinfo {author} {\bibfnamefont
  {I.}~\bibnamefont {Favero}}, \bibinfo {author} {\bibfnamefont
  {G.}~\bibnamefont {Leo}}, \ and\ \bibinfo {author} {\bibfnamefont
  {S.}~\bibnamefont {Ducci}},\ }\href {\doibase 10.1103/PhysRevLett.112.183901}
  {\bibfield  {journal} {\bibinfo  {journal} {Phys. Rev. Lett.}\ }\textbf
  {\bibinfo {volume} {112}},\ \bibinfo {pages} {183901} (\bibinfo {year}
  {2014})}\BibitemShut {NoStop}%
\bibitem [{\citenamefont {Horn}\ \emph {et~al.}(2012)\citenamefont {Horn},
  \citenamefont {Abolghasem}, \citenamefont {Bijlani}, \citenamefont {Kang},
  \citenamefont {Helmy},\ and\ \citenamefont {Weihs}}]{horn_12}%
  \BibitemOpen
  \bibfield  {author} {\bibinfo {author} {\bibfnamefont {R.}~\bibnamefont
  {Horn}}, \bibinfo {author} {\bibfnamefont {P.}~\bibnamefont {Abolghasem}},
  \bibinfo {author} {\bibfnamefont {B.~J.}\ \bibnamefont {Bijlani}}, \bibinfo
  {author} {\bibfnamefont {D.}~\bibnamefont {Kang}}, \bibinfo {author}
  {\bibfnamefont {A.~S.}\ \bibnamefont {Helmy}}, \ and\ \bibinfo {author}
  {\bibfnamefont {G.}~\bibnamefont {Weihs}},\ }\href {\doibase
  10.1103/PhysRevLett.108.153605} {\bibfield  {journal} {\bibinfo  {journal}
  {Phys. Rev. Lett.}\ }\textbf {\bibinfo {volume} {108}},\ \bibinfo {pages}
  {153605} (\bibinfo {year} {2012})}\BibitemShut {NoStop}%
\bibitem [{\citenamefont {Grassani}\ \emph {et~al.}(2015)\citenamefont
  {Grassani}, \citenamefont {Azzini}, \citenamefont {Liscidini}, \citenamefont
  {Galli}, \citenamefont {Strain}, \citenamefont {Sorel}, \citenamefont
  {Sipe},\ and\ \citenamefont {Bajoni}}]{Grassani_15}%
  \BibitemOpen
  \bibfield  {author} {\bibinfo {author} {\bibfnamefont {D.}~\bibnamefont
  {Grassani}}, \bibinfo {author} {\bibfnamefont {S.}~\bibnamefont {Azzini}},
  \bibinfo {author} {\bibfnamefont {M.}~\bibnamefont {Liscidini}}, \bibinfo
  {author} {\bibfnamefont {M.}~\bibnamefont {Galli}}, \bibinfo {author}
  {\bibfnamefont {M.~J.}\ \bibnamefont {Strain}}, \bibinfo {author}
  {\bibfnamefont {M.}~\bibnamefont {Sorel}}, \bibinfo {author} {\bibfnamefont
  {J.~E.}\ \bibnamefont {Sipe}}, \ and\ \bibinfo {author} {\bibfnamefont
  {D.}~\bibnamefont {Bajoni}},\ }\href {\doibase 10.1364/OPTICA.2.000088}
  {\bibfield  {journal} {\bibinfo  {journal} {Optica}\ }\textbf {\bibinfo
  {volume} {2}},\ \bibinfo {pages} {88} (\bibinfo {year} {2015})}\BibitemShut
  {NoStop}%
\bibitem [{\citenamefont {Silverstone}\ \emph {et~al.}(2015)\citenamefont
  {Silverstone}, \citenamefont {Santagati}, \citenamefont {Bonneau},
  \citenamefont {Strain}, \citenamefont {Sorel}, \citenamefont {O'Brien},\ and\
  \citenamefont {Thompson}}]{Silverstone_15}%
  \BibitemOpen
  \bibfield  {author} {\bibinfo {author} {\bibfnamefont {J.~W.}\ \bibnamefont
  {Silverstone}}, \bibinfo {author} {\bibfnamefont {R.}~\bibnamefont
  {Santagati}}, \bibinfo {author} {\bibfnamefont {D.}~\bibnamefont {Bonneau}},
  \bibinfo {author} {\bibfnamefont {M.~J.}\ \bibnamefont {Strain}}, \bibinfo
  {author} {\bibfnamefont {M.}~\bibnamefont {Sorel}}, \bibinfo {author}
  {\bibfnamefont {J.~L.}\ \bibnamefont {O'Brien}}, \ and\ \bibinfo {author}
  {\bibfnamefont {M.~G.}\ \bibnamefont {Thompson}},\ }\href
  {http://dx.doi.org/10.1038/ncomms8948} {\bibfield  {journal} {\bibinfo
  {journal} {Nature Communications}\ }\textbf {\bibinfo {volume} {6}},\
  \bibinfo {pages} {7948 EP } (\bibinfo {year} {2015})}\BibitemShut {NoStop}%
\bibitem [{\citenamefont {Chen}\ \emph {et~al.}(2016)\citenamefont {Chen},
  \citenamefont {Zhang}, \citenamefont {Zopf}, \citenamefont {Jung},
  \citenamefont {Zhang}, \citenamefont {Keil}, \citenamefont {Ding},\ and\
  \citenamefont {Schmidt}}]{chen_16}%
  \BibitemOpen
  \bibfield  {author} {\bibinfo {author} {\bibfnamefont {Y.}~\bibnamefont
  {Chen}}, \bibinfo {author} {\bibfnamefont {J.}~\bibnamefont {Zhang}},
  \bibinfo {author} {\bibfnamefont {M.}~\bibnamefont {Zopf}}, \bibinfo {author}
  {\bibfnamefont {K.}~\bibnamefont {Jung}}, \bibinfo {author} {\bibfnamefont
  {Y.}~\bibnamefont {Zhang}}, \bibinfo {author} {\bibfnamefont
  {R.}~\bibnamefont {Keil}}, \bibinfo {author} {\bibfnamefont {F.}~\bibnamefont
  {Ding}}, \ and\ \bibinfo {author} {\bibfnamefont {O.~G.}\ \bibnamefont
  {Schmidt}},\ }\href {http://dx.doi.org/10.1038/ncomms10387} {\bibfield
  {journal} {\bibinfo  {journal} {Nature Communications}\ }\textbf {\bibinfo
  {volume} {7}},\ \bibinfo {pages} {10387 EP } (\bibinfo {year}
  {2016})}\BibitemShut {NoStop}%
\bibitem [{\citenamefont {Atzeni}\ \emph {et~al.}(2018)\citenamefont {Atzeni},
  \citenamefont {Rab}, \citenamefont {Corrielli}, \citenamefont {Polino},
  \citenamefont {Valeri}, \citenamefont {Mataloni}, \citenamefont {Spagnolo},
  \citenamefont {Crespi}, \citenamefont {Sciarrino},\ and\ \citenamefont
  {Osellame}}]{atzeni2018integrated}%
  \BibitemOpen
  \bibfield  {author} {\bibinfo {author} {\bibfnamefont {S.}~\bibnamefont
  {Atzeni}}, \bibinfo {author} {\bibfnamefont {A.~S.}\ \bibnamefont {Rab}},
  \bibinfo {author} {\bibfnamefont {G.}~\bibnamefont {Corrielli}}, \bibinfo
  {author} {\bibfnamefont {E.}~\bibnamefont {Polino}}, \bibinfo {author}
  {\bibfnamefont {M.}~\bibnamefont {Valeri}}, \bibinfo {author} {\bibfnamefont
  {P.}~\bibnamefont {Mataloni}}, \bibinfo {author} {\bibfnamefont
  {N.}~\bibnamefont {Spagnolo}}, \bibinfo {author} {\bibfnamefont
  {A.}~\bibnamefont {Crespi}}, \bibinfo {author} {\bibfnamefont
  {F.}~\bibnamefont {Sciarrino}}, \ and\ \bibinfo {author} {\bibfnamefont
  {R.}~\bibnamefont {Osellame}},\ }\href@noop {} {\bibfield  {journal}
  {\bibinfo  {journal} {Optica}\ }\textbf {\bibinfo {volume} {5}},\ \bibinfo
  {pages} {311} (\bibinfo {year} {2018})}\BibitemShut {NoStop}%
\bibitem [{\citenamefont {Liscidini}\ and\ \citenamefont
  {Sipe}(2013)}]{liscidini_13}%
  \BibitemOpen
  \bibfield  {author} {\bibinfo {author} {\bibfnamefont {M.}~\bibnamefont
  {Liscidini}}\ and\ \bibinfo {author} {\bibfnamefont {J.~E.}\ \bibnamefont
  {Sipe}},\ }\href {\doibase 10.1103/PhysRevLett.111.193602} {\bibfield
  {journal} {\bibinfo  {journal} {Phys. Rev. Lett.}\ }\textbf {\bibinfo
  {volume} {111}},\ \bibinfo {pages} {193602} (\bibinfo {year}
  {2013})}\BibitemShut {NoStop}%
\bibitem [{\citenamefont {Rozema}\ \emph {et~al.}(2015)\citenamefont {Rozema},
  \citenamefont {Wang}, \citenamefont {Mahler}, \citenamefont {Hayat},
  \citenamefont {Steinberg}, \citenamefont {Sipe},\ and\ \citenamefont
  {Liscidini}}]{Rozema:15}%
  \BibitemOpen
  \bibfield  {author} {\bibinfo {author} {\bibfnamefont {L.~A.}\ \bibnamefont
  {Rozema}}, \bibinfo {author} {\bibfnamefont {C.}~\bibnamefont {Wang}},
  \bibinfo {author} {\bibfnamefont {D.~H.}\ \bibnamefont {Mahler}}, \bibinfo
  {author} {\bibfnamefont {A.}~\bibnamefont {Hayat}}, \bibinfo {author}
  {\bibfnamefont {A.~M.}\ \bibnamefont {Steinberg}}, \bibinfo {author}
  {\bibfnamefont {J.~E.}\ \bibnamefont {Sipe}}, \ and\ \bibinfo {author}
  {\bibfnamefont {M.}~\bibnamefont {Liscidini}},\ }\href {\doibase
  10.1364/OPTICA.2.000430} {\bibfield  {journal} {\bibinfo  {journal} {Optica}\
  }\textbf {\bibinfo {volume} {2}},\ \bibinfo {pages} {430} (\bibinfo {year}
  {2015})}\BibitemShut {NoStop}%
\bibitem [{\citenamefont {Fang}\ \emph {et~al.}(2016)\citenamefont {Fang},
  \citenamefont {Liscidini}, \citenamefont {Sipe},\ and\ \citenamefont
  {Lorenz}}]{Fang:16}%
  \BibitemOpen
  \bibfield  {author} {\bibinfo {author} {\bibfnamefont {B.}~\bibnamefont
  {Fang}}, \bibinfo {author} {\bibfnamefont {M.}~\bibnamefont {Liscidini}},
  \bibinfo {author} {\bibfnamefont {J.~E.}\ \bibnamefont {Sipe}}, \ and\
  \bibinfo {author} {\bibfnamefont {V.~O.}\ \bibnamefont {Lorenz}},\ }\href
  {\doibase 10.1364/OE.24.010013} {\bibfield  {journal} {\bibinfo  {journal}
  {Opt. Express}\ }\textbf {\bibinfo {volume} {24}},\ \bibinfo {pages} {10013}
  (\bibinfo {year} {2016})}\BibitemShut {NoStop}%
\bibitem [{\citenamefont {Crespi}\ \emph {et~al.}(2011)\citenamefont {Crespi},
  \citenamefont {Ramponi}, \citenamefont {Osellame}, \citenamefont {Sansoni},
  \citenamefont {Bongioanni}, \citenamefont {Sciarrino}, \citenamefont
  {Vallone},\ and\ \citenamefont {Mataloni}}]{crespi2011integrated}%
  \BibitemOpen
  \bibfield  {author} {\bibinfo {author} {\bibfnamefont {A.}~\bibnamefont
  {Crespi}}, \bibinfo {author} {\bibfnamefont {R.}~\bibnamefont {Ramponi}},
  \bibinfo {author} {\bibfnamefont {R.}~\bibnamefont {Osellame}}, \bibinfo
  {author} {\bibfnamefont {L.}~\bibnamefont {Sansoni}}, \bibinfo {author}
  {\bibfnamefont {I.}~\bibnamefont {Bongioanni}}, \bibinfo {author}
  {\bibfnamefont {F.}~\bibnamefont {Sciarrino}}, \bibinfo {author}
  {\bibfnamefont {G.}~\bibnamefont {Vallone}}, \ and\ \bibinfo {author}
  {\bibfnamefont {P.}~\bibnamefont {Mataloni}},\ }\href@noop {} {\bibfield
  {journal} {\bibinfo  {journal} {Nature communications}\ }\textbf {\bibinfo
  {volume} {2}},\ \bibinfo {pages} {566} (\bibinfo {year} {2011})}\BibitemShut
  {NoStop}%
\bibitem [{\citenamefont {Corrielli}\ \emph {et~al.}(2014)\citenamefont
  {Corrielli}, \citenamefont {Crespi}, \citenamefont {Geremia}, \citenamefont
  {Ramponi}, \citenamefont {Sansoni}, \citenamefont {Santinelli}, \citenamefont
  {Mataloni}, \citenamefont {Sciarrino},\ and\ \citenamefont
  {Osellame}}]{corrielli2014rotated}%
  \BibitemOpen
  \bibfield  {author} {\bibinfo {author} {\bibfnamefont {G.}~\bibnamefont
  {Corrielli}}, \bibinfo {author} {\bibfnamefont {A.}~\bibnamefont {Crespi}},
  \bibinfo {author} {\bibfnamefont {R.}~\bibnamefont {Geremia}}, \bibinfo
  {author} {\bibfnamefont {R.}~\bibnamefont {Ramponi}}, \bibinfo {author}
  {\bibfnamefont {L.}~\bibnamefont {Sansoni}}, \bibinfo {author} {\bibfnamefont
  {A.}~\bibnamefont {Santinelli}}, \bibinfo {author} {\bibfnamefont
  {P.}~\bibnamefont {Mataloni}}, \bibinfo {author} {\bibfnamefont
  {F.}~\bibnamefont {Sciarrino}}, \ and\ \bibinfo {author} {\bibfnamefont
  {R.}~\bibnamefont {Osellame}},\ }\href@noop {} {\bibfield  {journal}
  {\bibinfo  {journal} {Nature communications}\ }\textbf {\bibinfo {volume}
  {5}},\ \bibinfo {pages} {4249} (\bibinfo {year} {2014})}\BibitemShut
  {NoStop}%
\bibitem [{\citenamefont {Silverstone}\ \emph {et~al.}(2013)\citenamefont
  {Silverstone}, \citenamefont {Bonneau}, \citenamefont {Ohira}, \citenamefont
  {Suzuki}, \citenamefont {Yoshida}, \citenamefont {Iizuka}, \citenamefont
  {Ezaki}, \citenamefont {Natarajan}, \citenamefont {Tanner}, \citenamefont
  {Hadfield}, \citenamefont {Zwiller}, \citenamefont {Marshall}, \citenamefont
  {Rarity}, \citenamefont {O'Brien},\ and\ \citenamefont
  {Thompson}}]{silverstone_13}%
  \BibitemOpen
  \bibfield  {author} {\bibinfo {author} {\bibfnamefont {J.~W.}\ \bibnamefont
  {Silverstone}}, \bibinfo {author} {\bibfnamefont {D.}~\bibnamefont
  {Bonneau}}, \bibinfo {author} {\bibfnamefont {K.}~\bibnamefont {Ohira}},
  \bibinfo {author} {\bibfnamefont {N.}~\bibnamefont {Suzuki}}, \bibinfo
  {author} {\bibfnamefont {H.}~\bibnamefont {Yoshida}}, \bibinfo {author}
  {\bibfnamefont {N.}~\bibnamefont {Iizuka}}, \bibinfo {author} {\bibfnamefont
  {M.}~\bibnamefont {Ezaki}}, \bibinfo {author} {\bibfnamefont {C.~M.}\
  \bibnamefont {Natarajan}}, \bibinfo {author} {\bibfnamefont {M.~G.}\
  \bibnamefont {Tanner}}, \bibinfo {author} {\bibfnamefont {R.~H.}\
  \bibnamefont {Hadfield}}, \bibinfo {author} {\bibfnamefont {V.}~\bibnamefont
  {Zwiller}}, \bibinfo {author} {\bibfnamefont {G.~D.}\ \bibnamefont
  {Marshall}}, \bibinfo {author} {\bibfnamefont {J.~G.}\ \bibnamefont
  {Rarity}}, \bibinfo {author} {\bibfnamefont {J.~L.}\ \bibnamefont {O'Brien}},
  \ and\ \bibinfo {author} {\bibfnamefont {M.~G.}\ \bibnamefont {Thompson}},\
  }\href {http://dx.doi.org/10.1038/nphoton.2013.339} {\bibfield  {journal}
  {\bibinfo  {journal} {Nature Photonics}\ }\textbf {\bibinfo {volume} {8}},\
  \bibinfo {pages} {104 EP } (\bibinfo {year} {2013})}\BibitemShut {NoStop}%
\bibitem [{\citenamefont {Wang}\ \emph {et~al.}(2018)\citenamefont {Wang},
  \citenamefont {Paesani}, \citenamefont {Ding}, \citenamefont {Santagati},
  \citenamefont {Skrzypczyk}, \citenamefont {Salavrakos}, \citenamefont {Tura},
  \citenamefont {Augusiak}, \citenamefont {Man{\v c}inska}, \citenamefont
  {Bacco}, \citenamefont {Bonneau}, \citenamefont {Silverstone}, \citenamefont
  {Gong}, \citenamefont {Ac{\'\i}n}, \citenamefont {Rottwitt}, \citenamefont
  {Oxenl{\o}we}, \citenamefont {O{\textquoteright}Brien}, \citenamefont
  {Laing},\ and\ \citenamefont {Thompson}}]{Wangeaar7053}%
  \BibitemOpen
  \bibfield  {author} {\bibinfo {author} {\bibfnamefont {J.}~\bibnamefont
  {Wang}}, \bibinfo {author} {\bibfnamefont {S.}~\bibnamefont {Paesani}},
  \bibinfo {author} {\bibfnamefont {Y.}~\bibnamefont {Ding}}, \bibinfo {author}
  {\bibfnamefont {R.}~\bibnamefont {Santagati}}, \bibinfo {author}
  {\bibfnamefont {P.}~\bibnamefont {Skrzypczyk}}, \bibinfo {author}
  {\bibfnamefont {A.}~\bibnamefont {Salavrakos}}, \bibinfo {author}
  {\bibfnamefont {J.}~\bibnamefont {Tura}}, \bibinfo {author} {\bibfnamefont
  {R.}~\bibnamefont {Augusiak}}, \bibinfo {author} {\bibfnamefont
  {L.}~\bibnamefont {Man{\v c}inska}}, \bibinfo {author} {\bibfnamefont
  {D.}~\bibnamefont {Bacco}}, \bibinfo {author} {\bibfnamefont
  {D.}~\bibnamefont {Bonneau}}, \bibinfo {author} {\bibfnamefont {J.~W.}\
  \bibnamefont {Silverstone}}, \bibinfo {author} {\bibfnamefont
  {Q.}~\bibnamefont {Gong}}, \bibinfo {author} {\bibfnamefont {A.}~\bibnamefont
  {Ac{\'\i}n}}, \bibinfo {author} {\bibfnamefont {K.}~\bibnamefont {Rottwitt}},
  \bibinfo {author} {\bibfnamefont {L.~K.}\ \bibnamefont {Oxenl{\o}we}},
  \bibinfo {author} {\bibfnamefont {J.~L.}\ \bibnamefont
  {O{\textquoteright}Brien}}, \bibinfo {author} {\bibfnamefont
  {A.}~\bibnamefont {Laing}}, \ and\ \bibinfo {author} {\bibfnamefont {M.~G.}\
  \bibnamefont {Thompson}},\ }\href {\doibase 10.1126/science.aar7053}
  {\bibfield  {journal} {\bibinfo  {journal} {Science}\ } (\bibinfo {year}
  {2018}),\ 10.1126/science.aar7053},\ \Eprint
  {http://arxiv.org/abs/http://science.sciencemag.org/content/early/2018/03/07/science.aar7053.full.pdf}
  {http://science.sciencemag.org/content/early/2018/03/07/science.aar7053.full.pdf}
  \BibitemShut {NoStop}%
\bibitem [{\citenamefont {Eckstein}\ \emph {et~al.}(2014)\citenamefont
  {Eckstein}, \citenamefont {Boucher}, \citenamefont {Lemaître}, \citenamefont
  {Filloux}, \citenamefont {Favero}, \citenamefont {Leo}, \citenamefont {Sipe},
  \citenamefont {Liscidini},\ and\ \citenamefont {Ducci}}]{Eckstein_14}%
  \BibitemOpen
  \bibfield  {author} {\bibinfo {author} {\bibfnamefont {A.}~\bibnamefont
  {Eckstein}}, \bibinfo {author} {\bibfnamefont {G.}~\bibnamefont {Boucher}},
  \bibinfo {author} {\bibfnamefont {A.}~\bibnamefont {Lemaître}}, \bibinfo
  {author} {\bibfnamefont {P.}~\bibnamefont {Filloux}}, \bibinfo {author}
  {\bibfnamefont {I.}~\bibnamefont {Favero}}, \bibinfo {author} {\bibfnamefont
  {G.}~\bibnamefont {Leo}}, \bibinfo {author} {\bibfnamefont {J.~E.}\
  \bibnamefont {Sipe}}, \bibinfo {author} {\bibfnamefont {M.}~\bibnamefont
  {Liscidini}}, \ and\ \bibinfo {author} {\bibfnamefont {S.}~\bibnamefont
  {Ducci}},\ }\href {\doibase 10.1002/lpor.201400057} {\bibfield  {journal}
  {\bibinfo  {journal} {Laser \& Photonics Reviews}\ }\textbf {\bibinfo
  {volume} {8}},\ \bibinfo {pages} {L76} (\bibinfo {year} {2014})}\BibitemShut
  {NoStop}%
\bibitem [{\citenamefont {Fang}\ \emph {et~al.}(2014)\citenamefont {Fang},
  \citenamefont {Cohen}, \citenamefont {Liscidini}, \citenamefont {Sipe},\ and\
  \citenamefont {Lorenz}}]{Fang:14}%
  \BibitemOpen
  \bibfield  {author} {\bibinfo {author} {\bibfnamefont {B.}~\bibnamefont
  {Fang}}, \bibinfo {author} {\bibfnamefont {O.}~\bibnamefont {Cohen}},
  \bibinfo {author} {\bibfnamefont {M.}~\bibnamefont {Liscidini}}, \bibinfo
  {author} {\bibfnamefont {J.~E.}\ \bibnamefont {Sipe}}, \ and\ \bibinfo
  {author} {\bibfnamefont {V.~O.}\ \bibnamefont {Lorenz}},\ }\href {\doibase
  10.1364/OPTICA.1.000281} {\bibfield  {journal} {\bibinfo  {journal} {Optica}\
  }\textbf {\bibinfo {volume} {1}},\ \bibinfo {pages} {281} (\bibinfo {year}
  {2014})}\BibitemShut {NoStop}%
\bibitem [{\citenamefont {Jizan}\ \emph {et~al.}(2015)\citenamefont {Jizan},
  \citenamefont {Helt}, \citenamefont {Xiong}, \citenamefont {Collins},
  \citenamefont {Choi}, \citenamefont {Joon~Chae}, \citenamefont {Liscidini},
  \citenamefont {Steel}, \citenamefont {Eggleton},\ and\ \citenamefont
  {Clark}}]{Iman_15}%
  \BibitemOpen
  \bibfield  {author} {\bibinfo {author} {\bibfnamefont {I.}~\bibnamefont
  {Jizan}}, \bibinfo {author} {\bibfnamefont {L.~G.}\ \bibnamefont {Helt}},
  \bibinfo {author} {\bibfnamefont {C.}~\bibnamefont {Xiong}}, \bibinfo
  {author} {\bibfnamefont {M.~J.}\ \bibnamefont {Collins}}, \bibinfo {author}
  {\bibfnamefont {D.-Y.}\ \bibnamefont {Choi}}, \bibinfo {author}
  {\bibfnamefont {C.}~\bibnamefont {Joon~Chae}}, \bibinfo {author}
  {\bibfnamefont {M.}~\bibnamefont {Liscidini}}, \bibinfo {author}
  {\bibfnamefont {M.~J.}\ \bibnamefont {Steel}}, \bibinfo {author}
  {\bibfnamefont {B.~J.}\ \bibnamefont {Eggleton}}, \ and\ \bibinfo {author}
  {\bibfnamefont {A.~S.}\ \bibnamefont {Clark}},\ }\href@noop {} {\bibfield
  {journal} {\bibinfo  {journal} {Scientific Reports}\ }\textbf {\bibinfo
  {volume} {5}},\ \bibinfo {pages} {12557 EP } (\bibinfo {year}
  {2015})}\BibitemShut {NoStop}%
\bibitem [{\citenamefont {Grassani}\ \emph {et~al.}(2016)\citenamefont
  {Grassani}, \citenamefont {Simbula}, \citenamefont {Pirotta}, \citenamefont
  {Galli}, \citenamefont {Menotti}, \citenamefont {Harris}, \citenamefont
  {Baehr-Jones}, \citenamefont {Hochberg}, \citenamefont {Galland},
  \citenamefont {Liscidini},\ and\ \citenamefont {Bajoni}}]{Grassani_16}%
  \BibitemOpen
  \bibfield  {author} {\bibinfo {author} {\bibfnamefont {D.}~\bibnamefont
  {Grassani}}, \bibinfo {author} {\bibfnamefont {A.}~\bibnamefont {Simbula}},
  \bibinfo {author} {\bibfnamefont {S.}~\bibnamefont {Pirotta}}, \bibinfo
  {author} {\bibfnamefont {M.}~\bibnamefont {Galli}}, \bibinfo {author}
  {\bibfnamefont {M.}~\bibnamefont {Menotti}}, \bibinfo {author} {\bibfnamefont
  {N.~C.}\ \bibnamefont {Harris}}, \bibinfo {author} {\bibfnamefont
  {T.}~\bibnamefont {Baehr-Jones}}, \bibinfo {author} {\bibfnamefont
  {M.}~\bibnamefont {Hochberg}}, \bibinfo {author} {\bibfnamefont
  {C.}~\bibnamefont {Galland}}, \bibinfo {author} {\bibfnamefont
  {M.}~\bibnamefont {Liscidini}}, \ and\ \bibinfo {author} {\bibfnamefont
  {D.}~\bibnamefont {Bajoni}},\ }\href {http://dx.doi.org/10.1038/srep23564}
  {\bibfield  {journal} {\bibinfo  {journal} {Scientific Reports}\ }\textbf
  {\bibinfo {volume} {6}},\ \bibinfo {pages} {23564 EP } (\bibinfo {year}
  {2016})}\BibitemShut {NoStop}%
\bibitem [{\citenamefont {Barbieri}\ \emph {et~al.}(2005)\citenamefont
  {Barbieri}, \citenamefont {Cinelli}, \citenamefont {Mataloni},\ and\
  \citenamefont {De~Martini}}]{mata:hyperrealization}%
  \BibitemOpen
  \bibfield  {author} {\bibinfo {author} {\bibfnamefont {M.}~\bibnamefont
  {Barbieri}}, \bibinfo {author} {\bibfnamefont {C.}~\bibnamefont {Cinelli}},
  \bibinfo {author} {\bibfnamefont {P.}~\bibnamefont {Mataloni}}, \ and\
  \bibinfo {author} {\bibfnamefont {F.}~\bibnamefont {De~Martini}},\
  }\href@noop {} {\bibfield  {journal} {\bibinfo  {journal} {Physical Review
  A}\ }\textbf {\bibinfo {volume} {72}},\ \bibinfo {pages} {052110} (\bibinfo
  {year} {2005})}\BibitemShut {NoStop}%
\bibitem [{\citenamefont {James}\ \emph {et~al.}(2001)\citenamefont {James},
  \citenamefont {Kwiat}, \citenamefont {Munro},\ and\ \citenamefont
  {White}}]{kwiat:numericoptimization}%
  \BibitemOpen
  \bibfield  {author} {\bibinfo {author} {\bibfnamefont {D.~F.}\ \bibnamefont
  {James}}, \bibinfo {author} {\bibfnamefont {P.~G.}\ \bibnamefont {Kwiat}},
  \bibinfo {author} {\bibfnamefont {W.~J.}\ \bibnamefont {Munro}}, \ and\
  \bibinfo {author} {\bibfnamefont {A.~G.}\ \bibnamefont {White}},\ }\href@noop
  {} {\bibfield  {journal} {\bibinfo  {journal} {Physical Review A}\ }\textbf
  {\bibinfo {volume} {64}} (\bibinfo {year} {2001})}\BibitemShut {NoStop}%
\bibitem [{\citenamefont {Altepeter}\ \emph
  {et~al.}(2005{\natexlab{b}})\citenamefont {Altepeter}, \citenamefont
  {Jeffrey},\ and\ \citenamefont {Kwiat}}]{Altepeter:05}%
  \BibitemOpen
  \bibfield  {author} {\bibinfo {author} {\bibfnamefont {J.~B.}\ \bibnamefont
  {Altepeter}}, \bibinfo {author} {\bibfnamefont {E.~R.}\ \bibnamefont
  {Jeffrey}}, \ and\ \bibinfo {author} {\bibfnamefont {P.~G.}\ \bibnamefont
  {Kwiat}},\ }\href {\doibase 10.1364/OPEX.13.008951} {\bibfield  {journal}
  {\bibinfo  {journal} {Opt. Express}\ }\textbf {\bibinfo {volume} {13}},\
  \bibinfo {pages} {8951} (\bibinfo {year} {2005}{\natexlab{b}})}\BibitemShut
  {NoStop}%
\end{thebibliography}
\end{document}